\documentclass[11pt,letterpaper]{article} 

\usepackage[]{amsthm}
\usepackage[]{amsmath}
\usepackage[]{amssymb}
\usepackage[]{bbm}
\usepackage[noend, ruled, linesnumbered, noline, longend]{algorithm2e}

\DeclareMathOperator*{\argmin}{arg\,min}
\SetKwInOut{Parameter}{Parameters}
\usepackage[dvipsnames]{xcolor}  % textcolor command \definecolor{mintgreen}{rgb}{0.0, 0.47, 0.44}

\usepackage[]{hyperref}
\usepackage{hyphenat}
\usepackage{tikz}
\usepackage{subfig}
\usepackage{graphicx}

\tikzset{%
    every node/.style={draw,circle,minimum size=1cm},
    labelnode/.style={draw=none,rectangle,minimum size=0cm},
    garc/.style={thick,-latex},
    warc/.style={thick,-latex}
}

\usepackage{easyReview}

\usepackage{booktabs}

\usepackage{natbib}
\usepackage[capitalise]{cleveref} 
\usepackage{apalike}

\newtheorem{definition}{Definition}[section]
\newtheorem{proposition}{Proposition}[section]
\newtheorem{lemma}[proposition]{Lemma}

\newtheorem{corollary}[proposition]{Corollary}
\theoremstyle{definition}
\newtheorem{example}{Example}[section]

\newtheorem{rem}{Remark}[section]

\usepackage{mathtools}
\usepackage{centernot}
\newcommand{\POSP}{\text{POSP}}          % Partial Order Shortest Path Problem
\newcommand{\SP}{SP}              % Shortest Path Problem
\newcommand{\POSPMAX}{\text{\POSP}_{max}}% Partial Order Shortest Path Problem (max complete set)
\newcommand{\POSPMIN}{\text{\POSP}_{min}}% Partial Order Shortest Path Problem (min complete set)
\newcommand{\MOSP}{MOSP}          % Multi Objective Shortest Path Problem

\newcommand{\strictpo}{\prec}   % Strict Partial Order Operator
\newcommand{\strictNpo}{\nprec} % Negated Strict Partial Order Operator
\newcommand{\strictsuccpo}{\succ}

\newcommand{\po}{\preceq}       % Partial Order Operator
\newcommand{\succpo}{\succeq}   % Partial Order Operator
\newcommand{\npo}{\npreceq}     % Negated Partial Order Operator

\newcommand{\LEO}[1][\po]{\mathbb{O}_{#1}}   % Linear Extension Oracle 
\newcommand{\strictto}{\vartriangleleft}              % Strict Total Order Operator
\newcommand{\TO}{\trianglelefteq}                  % Total Order Operator
\newcommand{\chain}{\mathcal{C}}  % a chain in the partial ordered valuation set
\newcommand{\eff}{\mathcal{E}}    % Set of All Efficient Walks
\newcommand{\effopt}{\eff^*}      % Minimal Complete Set of Efficient Walks
\newcommand{\algeff}{\eff}        % Sets Calculated by the Algorithms
\newcommand{\nondom}{\mathcal{N}} % Nondominated Points: Pareto Front

\newcommand{\VS}{\mathcal{T}}                    % Weight Set
\newcommand{\PVS}{(\VS,\po)}                     % Partial Order
\newcommand{\va}{\tau}                           % Valuation
\newcommand{\valt}{\tau'}                        % Some Other Valuation
\newcommand{\vao}{\tau^*}                        % Some 'Optimal' Valuation
\newcommand{\wid}{\omega}                        % Width of the partially ordered set

\newcommand{\la}{l}                            % Some Label
\newcommand{\join}{\lor}          % Join Operator
\newcommand{\G}{G}                         % Graph
\newcommand{\A}{A}                         % Arc Set
\newcommand{\V}{V}                         % Vertex Set
\newcommand{\vrt}{v}                       % Vertex
\newcommand{\arc}{a}                       % Arc
\newcommand{\vrtalt}{u}                    % Another Vertex
\newcommand{\WW}{c}                        % Walk Weights
\newcommand{\SPS}{\mathcal{S}}              % Simple Paths of a Graph
\newcommand{\PS}{\mathcal{P}}              % Walks of a Graph
\newcommand{\p}{p}                         % Single Path
\newcommand{\palt}{p'}                     % Another Path
\newcommand{\paalt}{p''}                   % Another Path
\newcommand{\UF}[1][\WW]{\mathcal{U}_{#1}} % Arc Update function
\newcommand{\outgoing}[1]{\delta^{+}(#1)}  % out neighborhood
\newcommand{\ingoing}[1]{\delta^{-}(#1)}  % out neighborhood

\newcommand{\N}{\mathbb{N}}       % Natural Numbers
\newcommand{\R}{\mathbb{R}}       % Real Numbers

\newcommand{\instance}{(\G,\WW,\VS,\po,s)}
\newcommand{\objSpace}{\R^d} % objective space

\newcommand{\ON}{\mathcal{O}}      % O-Notation
\newcommand{\ENC}{\Gamma}       % Unspecified Encoding length parameter of instance

\newcommand{\efflen}{\mu}           % Efficiency Length of \WW and \G

\newcommand{\Null}{\mathtt{NULL}}

\newcommand{\ob}{\textopenbullet}
\newcommand{\fb}{\textbullet}

\DeclareMathOperator{\image}{Im}

\DeclareMathOperator{\minmerge}{\mathrm{minmerge}}
\DeclareMathOperator{\maxmerge}{\mathrm{maxmerge}}

\newcommand\fundingfootnote[1]{%
  \let\thefootnote\relax%
  \footnotetext{\textit{#1}}%
  \let\thefootnote\svthefootnote%
}

\hypersetup{
	pdftitle={Labeling Methods for Partially Ordered Paths},
	pdfsubject={},
	pdfauthor={Ricardo Euler, Pedro Maristany de las Casas},
        pdfkeywords={Dynamic Programming, Partial Order, Shortest Path, Multi-Objective, Dijkstra},
}

\title{Labeling Methods for Partially Ordered Paths}
\date{}
\author{ \href{https://orcid.org/0000-0001-5112-4191}{\includegraphics[scale=0.06]{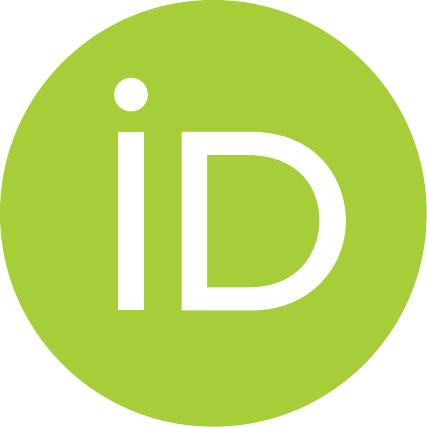}\hspace{1mm}Ricardo Euler}
        \href{https://orcid.org/0000-0002-4197-0893}{\includegraphics[scale=0.06]{orcid_id_icon.pdf}\hspace{1mm}Pedro Maristany de las Casas}\\
	Network Optimization \\
	Zuse Institute Berlin\\
	Germany, Berlin, 14195\\
	\texttt{\{euler,maristany\}@zib.de}\\ }

\usepackage{fancyhdr}
\begin{document}

\maketitle       

\begin{abstract}
The landscape of applications and subroutines relying on shortest path computations continues to grow steadily. This growth is driven by the undeniable success of shortest path algorithms in theory and practice. It also introduces new challenges as the models and assessing the optimality of paths become more complicated. Hence, multiple recent publications in the field adapt existing labeling methods in an ad hoc fashion to their specific problem variant without considering the underlying general structure: they always deal with multi-criteria scenarios, and those criteria define different partial orders on the paths. In this paper, we introduce the partial order shortest path problem (\POSP{}), a generalization of the multi-objective shortest path problem (\MOSP{}) and in turn also of the classical shortest path problem. \POSP{} captures the particular structure of many shortest path applications as special cases. In this generality, we study optimality conditions or the lack of them, depending on the objective functions' properties. Our final contribution is a big lookup table summarizing our findings and providing the reader with an easy way to choose among the most recent multi-criteria shortest path algorithms depending on their problems' weight structure. Examples range from  time-dependent shortest path and bottleneck path problems to the 
electric vehicle shortest path problem with recharging and complex financial weight functions studied in the public transportation community. Our results hold for general digraphs and, therefore, surpass previous generalizations that were limited to acyclic graphs.
\end{abstract}

\fundingfootnote{Funding: This work was supported by the Research Campus MODAL funded by the German Federal Ministry of Education and Research (BMBF) [grant number 05M20ZBM].}

\providecommand{\keywords}[1]{\noindent\textbf{\textit{Index terms---}} #1}
\providecommand{\competinginterest}[1]{\noindent\textbf{\textit{Declaration of interest---}} #1}


\competinginterest{none}

\section{Introduction} 

In the multi-objective shortest path problem (\MOSP{}), every arc of a directed graph $\G=(\V,\A)$ carries a vector-valued weight drawn from $\objSpace{}$; a path's weight is calculated by summing over its arcs.  \MOSP{} aims to find complete sets of efficient paths from a start vertex to all other vertices. Here, a path is called efficient and its weight nondominated if no other path has lower weight in one dimension and at least equivalent weight in all others.
Expressed in terms of order theory,  a path is efficient if it is minimal with respect to the canonical partial order on $\objSpace{}$, which is 
the $d$-fold product order of the \emph{lesser-or-equal} total order $\leq$ on $\R$.
A set of efficient paths is complete if it contains a representative for every nondominated weight.
\MOSP{} is often considered in the OR community \citep[e.g.,][]{Hansen1980, Martins1984, Boekler2018, MaristanydelasCasasSedenoNodaBorndoerfer2021, Sedenonoda2019} as well as more recently in the AI community \citep{Zhang2022,Hernandez2023,Ahmadi2022}.

In other path-finding applications, calculating a path's weight is more complex than in MOSP, and efficient paths are defined with respect to other partial orders.
For example, \cite{Euler2022} tackled the problem of finding minimum cost paths in public transportation networks. 
Here, a path's cost is calculated by complex business rules depending on traversed fare zones, time, and transfers.
The authors address this issue by introducing arc weights that are drawn from a partially ordered monoid and a digraph modeling public transit tickets.
This results in an intricate partial order on the paths' weight space.
By adapting the Multi-Objective RAPTOR algorithm \citep{Delling2015}, they compute complete efficient sets, from which the minimum cost path is then extracted in a post-processing step.

Other authors took to showing that the label-setting method by \cite{Martins1984} can be adapted to their applications.
Weights and weight functions that have been considered are, e.g., intervals \citep{Okada1994}, fuzzy numbers \citep{Okada2000}, belief functions \citep{Vu2022}, colors \citep{Ensor2011}, bottleneck weight functions \citep{Gandibleux2006} and battery charge functions \citep{Baum2019}.
While not always made explicit, all papers mentioned above study special cases of partial orders on the set of paths and rely on similar arguments.

\subsection{Contribution}

We introduce the \emph{partial order shortest path problem} (POSP). All applications mentioned so far are special cases of POSP. An instance of \POSP{} consists of a directed graph $G = (V,A)$ in which every path is assigned a weight drawn from a partially ordered \emph{weight space}. 

For a given source $s\in\V$, \POSP{} asks to find a set of efficient $s$-$v$-paths for every $\vrt \in \V$.
The formal definitions of efficiency and the output set are given in \Cref{def:POSP}. 

For \MOSP{} and for the bottleneck shortest path problem, label-setting and label-correcting algorithms are used in the literature. These algorithms build paths arc by arc. Given an $s$-$v$-path $p$ whose costs are known and an arc $a$ in the forward star of $v$, the costs of the concatenation of $p$ and $a$ are immediate in these scenarios. Only a sum or a $\min$ operation needs to be conducted. 
However, in \POSP{}, paths are assigned a weight without further specifications on how they are obtained from the paths' arcs. Such a cost update after a concatenation might not be trivial. 
The suitability of labeling algorithms in these scenarios is thus not guaranteed.
This begs the question: Given that all these applications are special cases of \POSP{}, what are the least restrictive conditions on the paths' costs structure such that label-setting and label-correcting algorithms can be used?
Our main contribution in the paper is the identification of such conditions. Based on them, we derive label-setting and label-correcting algorithms for \POSP{}. 

In addition to the introduction of \POSP{}, we contribute to an enhancement of the running time bounds in multiple of the aforementioned publications \citep[e.g.,][]{Okada1994,Okada2000,Gandibleux2006}. We achieve the improvement by choosing the recently introduced Multi-Objective Dijkstra Algorithm  (MDA) \citep{MaristanydelasCasasSedenoNodaBorndoerfer2021} for \MOSP{} as a basis for our label-setting \POSP{} method. As in the \MOSP{} framework, the MDA for \POSP{} improves the best output-sensitive running time bound known previously for algorithms for these applications.

The MDA for \POSP{} solves applications in which efficient paths may contain cycles. Thereby, we
derive a label-setting algorithm for the \emph{weight constrained shortest path problem with replenishment}, for which only a label-correcting method was known \citep{Smith2012}
and for a variant of the \emph{electric vehicle shortest path problem with (discretized) recharging}, for which cycles were  thus far resolved by network expansion \citep{Strehler2017}.

\Cref{tab:summary} is a summary of our results and an overview that facilitates choosing algorithms depending on the weight structure of the considered problem.

\subsection{Comparison to Literature}

In the following, we distinguish \POSP{} from various previous results in the literature.
The most similar results are due to \cite{Perny2005} who aim to find maximal cardinality complete efficient sets w.r.t to a quasi-transitive binary relation on the paths. Their approach is, however, limited in that it is suitable only for acyclic digraphs. It also cannot be used to find minimum cardinality complete sets of efficient paths, a problem variant markedly more common in practice.
We will show in \cref{subsec:labelsetting} that our approach can be used to compute efficient sets w.r.t to quasi-transitive binary relations on general digraphs.

The bottleneck weight function studied by \cite{Gandibleux2006} allows for efficient paths with dominated subpaths. This violates the principle of 
optimality at the heart of \MOSP{} algorithms.
The authors address this problem by adapting Martins' label-setting to construct supersets of maximal complete sets of efficient paths.
Hence, dominated solutions must be filtered out in an additional step, making the algorithm no longer label-setting in the strict sense.
We show that the bottleneck weight function fulfills a relaxed form of subpath optimality that allows the computation of minimal complete sets of efficient paths  using a label-setting algorithm without the need to post-process the com\-pu\-ted set of paths.

\cite{Reinhardt2011} consider an extension of \MOSP{} with several non-additive weight functions. Their framework, however, remains restricted to product orders and does not offer dominance rules for, e.g., path weights drawn from a subset lattice. Subset lattices are a common weight structure in practice as a model for, e.g., fare zones in public transportation. The \POSP{} framework offers natural dominance rules for all lattice-based weight structures.

\cite{Parmentier2019a} studies a generalization of the resource-constrained shortest path problem in which costs and weights can be drawn from lattice-ordered monoids. The optimality conditions that we identify for \POSP{} are more general as they are fulfilled by semilattices that are not monoids and partially ordered, positive monoids that are not lattices.

\cite{Carraway1990} study a generalization of \MOSP{} in which each arc carries a weight in $\R^d$ and a path's weight is computed via a binary relation on $\R$. They propose a generalized dynamic programming approach based on a \emph{weak principle of optimality} stating that ``an optimal path must be composed of subpaths that can be part of an optimal path''. This weak principle of optimality is, however, a tautology that is true for all kinds of path problems. It is hence not very useful in algorithm design.

Finally, note that \POSP{} is neither a special case of the algebraic path problem  \citep[cf.][]{fink1992survey} nor vice versa.
For every pair of vertices $s,t$, the algebraic path problem aims at calculating a weighted sum (w.r.t. to an arbitrary semiring) of all simple paths between $s$ and $t$.  
As a result, it is somewhat broader than \POSP{}, which always uses the minimum in lieu of the sum operator. However, it is also more limited in that the solution set only ever contains one path for each pair of vertices.

\subsection{Outline}
After introducing basic notation in the remainder of this chapter,
we formalize \POSP{} in \cref{sec:posp}.
In \cref{sec:dpp}, we develop conditions under which \POSP{} instances exhibit optimal substructure and analyze the relations among them.
Based on these conditions, we develop exemplary label-correcting and label-setting algorithms in \cref{sec:algorithms}. 
\cref{sec:algorithms} also includes an output-sensitivity analysis and a structured overview, matching \POSP{} variants with an appropriate algorithm.
\Cref{sec:conclusion} offers a conclusion.

\subsection{Notation}\label{sec:notation}

We denote consecutive sets of integers by the following  convention:
For 
$n\in\N$, let  
$[n] := \{1,\dots,n\}$ and $[n]_0 := \{0,\dots,n\}$.

Directed graphs are denoted by $\G=(\V,\A)$. They may contain loops, but not parallel arcs.
This restriction is made only to simplify notation; all results also hold for general digraphs.
A path in $\G$ is a sequence of vertices $(v_1,\dots,v_n)$ such that $(v_i,v_{i+1})\in \A$ for $i\in [n-1]$. $\PS$ denotes the set of all paths in $\G$.
This definition includes paths of the form $(v)$ for $\vrt\in \V$.
In a path, vertices may repeat. 
Paths that do not repeat vertices are called \emph{simple paths}. 
The set of simple paths in $\G$ is $\SPS \subseteq \PS$.
%A path  $\p \not \in \SPS$ is called a \emph{path}.
For two nodes $s,t \in V$, the set $\PS_{s,t}$ ($\SPS_{s,t}$) denotes the sets of $s,t$-paths (simple $s,t$-paths) in $\G$;
the set $\PS_{s}$ denotes all paths starting in $s\in\V$.
A path $\p\in\PS_{t,t}$ is called a \emph{cycle}. If it repeats no vertex but $t$, and $t$ appears exactly twice, it is called 
a simple cycle. 
The length of a path  $\p = (v_1,\dots,v_n) \in \PS$ is the number of its arcs and hence defined as $|\p| := n-1$.
For two paths $\p\in \PS_{s,v}$ and $\palt \in \PS_{v,t}$, we denote their concatenation by $\p+\palt$.
The same notation is used for appending an arc $\arc\in\outgoing{v}$ to $\p$, i.e., we write $\p+\arc$.

A partially ordered set ($S,\preceq)$
consists of  a ground set $S$ and a binary relation $\preceq$ on $S$ that fulfills
 $x \preceq y \land y \preceq z \implies x \preceq z \, \forall x,y,z \in S$ (transitivity),
 $x \preceq y \land y \preceq x \implies x=y          \, \forall x,y\in S$ (antisymmetry)
and $x\preceq x  \,   \forall x \in S$ (reflexivity).
If antisymmetry of $\preceq$ is not required, $(S,\preceq)$ is a preordered set.
A subset $\chain \subseteq S$ is called a \emph{chain} if $\po$ induces a total order on $\chain$.
Every binary relation $\preceq$ induces a strict binary relation $\prec$ given as
 $x \prec y :\Leftrightarrow x \preceq y \land x \neq y.$
Finally, we call $\preceq$ \emph{quasi-transitive} if $\prec$ is transitive but not $\preceq$ itself.

For a function $f: A\to B$ and a set $S\subseteq A$, we denote by $f(S)$ the image of $S$ under $f$.
In the case $S = A$, we also write $\image(f):= f(A)$.

For any binary relation $\po\, : B\times B \to \{True,False\}$, we write  $f(A)\po b$ to mean  $\exists b' \in f(A):  b' \po b$.
The negated expression $f(A)\ \npo b$ means $\forall b' \in f(A):  b' \npo b$.
For two functions $f: A \to B$ and $g: C \to D$ we denote their diagonal by $f\Delta g$, i.e.,
$f\Delta g: A\times C \to B \times D \, \, (a,c)  \mapsto \left(f(a),g(c)\right)$.

\section{The Partial Order Shortest Path Problem}\label{sec:posp}

An instance of \POSP{} is denoted by $\instance$,
where  $\G=(\V,\A)$ is a digraph, $\WW: \PS \to \VS$ is a weight function 
mapping from $\PS$ into a partial ordered weight space $(\VS,\po)$, and $s\in \V$ is the origin of the search.
Note that $\WW$ and the partial order $\po$ induce a preorder on $\PS_s$.
Following common notation \citep[cf.][]{Ehrgott2005}, 
we call a path  $\p \in \PS_{s,t}$ \emph{efficient} if there is no other $s,t$-path $\palt\in \PS_{s,t}$ with $\palt \strictpo \p$. 
The corresponding weight $\WW(\p)$ is called \emph{nondominated}.
We denote the set of all efficient \mbox{$s,t$-paths} by $\eff_{s,t}$.  
The set of all nondominated weights $\nondom_{s,t} := \WW(\eff_{s,t})$ is called the \emph{nondominated set}.
We call a set of $s,t$-paths $\effopt_{s,t}$ a \emph{complete set of efficient $s,t$-paths} if 
$\nondom_{s,t}  = \WW(\effopt_{s,t})$.
$\effopt_{s,t}$ is called a \emph{minimal (maximal) complete set of efficient $s,t$-paths} 
if there is no other complete set of efficient $s,t$-paths $\tilde{\mathcal{E}}_{s,t}$ with $\tilde{\mathcal{E}}_{s,t} \subset \effopt_{s,t}$ ($\tilde{\mathcal{E}}_{s,t} \supset \effopt_{s,t}$).
Equipped with the above notation, we can now state the \emph{partial order shortest path problem} (\POSP).

\begin{definition}[$\POSP$]\label{def:POSP}
Given a \POSP{} instance $\instance$,
the \emph{partial order shortest path problem} (\POSP) asks to find a complete set of efficient 
$s,v$-paths  $\effopt_{s,v}$ for all $\vrt\in\V$.
The problem of finding minimal (maximal) sets $\effopt_{s,v},v\in \V$ is denoted by  $\POSPMIN$ ($\POSPMAX$).
\end{definition}

We call a \POSP{} instance \emph{well-posed} if every chain in $\image(\WW)$ has a minimum element and $|\bigcup_{\vrt\in \V}\nondom_{s,\vrt}| < \infty$.
If an instance is not well-posed, the nondominated set is either of infinite size or there is an infinite sequence of paths with strictly decreasing weight. 
In the remainder of this paper, every \POSP{} instance considered is assumed to be well-posed. 

A special case of 
\POSP{} is the \emph{multi-objective shortest path problem}.
\begin{definition}[$\MOSP$]\label{def:mosp}
   A \POSP{} instance $\instance$ is a $d$-dimension\-al \MOSP{} instance if and only if  
   $(\VS,\po) \subseteq (\objSpace{},\leq)$, where $\leq$ is the canonical partial order on $\objSpace{}$,
   and there is a vector $ \mathbb{X} \in \R^{[d]\times \A }$ such that for all $v\in V$ and  $\p\in\PS_{s,v}$ it holds that 
   $
   \WW(\p + \arc ) = \WW(\p) + \mathbb{X}_{a} \, \forall \arc \in \outgoing{v}.
   $ 
   The corresponding problem is called \MOSP{}.
\end{definition}

An instance of \MOSP{} is well-posed if and only if there is no cycle with negative weight in any cost component.
Hence, (minimal) complete sets of efficient paths can be constructed from simple paths alone.
Well-posed \MOSP{} can be solved with label-correcting algorithms; to use a label-setting algorithm,
one additionally needs the arc weights $\mathbb{X}_a$ to be component\-wise nonnegative \citep{Ehrgott2005}.
This is not a real restriction: As in the single-criteria case \citep{Johnson1977}, 
any well-posed \MOSP{} can be transformed into one with  $\mathbb{X}_a \geq 0$ by building the multidimensional reduced costs \citep[][Section 2]{Sedenonoda2019}.
In the generality of \POSP{}, such a transformation is not necessarily possible.

We are interested in identifying conditions under which  
labeling methods
for \MOSP{} can be generalized to \POSP{}.
To this end, we generalize the notions of nonnegative cycles and arc weights to \POSP.

\begin{definition}
    We call a \POSP{} instance $\instance$ cycle-non-decreasing if $\forall v \in V \ \forall \p \in \PS_{s,v} \, \forall \palt \in \PS_{v,v} \backslash \{(v)\}: \WW(\p + \palt) \strictNpo \WW(\p)$
    and arc-non-decreasing if $\forall v \in V \, \forall \p \in \PS_{s,v} \, \forall \arc \in \outgoing{v}:   \WW(\p + \arc) \strictNpo \WW(\p)$.
    Replacing the operator $\strictNpo$ by $\succpo$, we define the terms
    cycle-increasing  and arc-increasing. Replacing it by $\strictsuccpo$ defines the terms strictly-cycle-increasing and strictly-arc-increasing.
\end{definition}

Arc-increasing instances are cycle-increasing; arc-non-decreasing instan\-ces are, however, not necessarily cycle-non-decreasing. Note also that even strict arc-increasingness does not imply well-posedness.

Finally, we introduce \emph{$\efflen$-bounded} \POSP{} instances.
If $\instance$ is $\efflen$-bounded, there is no efficient path longer than $\efflen$.
Note that this property is stricter than 
$|\bigcup_{\vrt\in \V}\nondom_{s,\vrt}| < \infty$.
$\POSPMAX{}$ instances must always be $\efflen$-bounded.
For $\text{\MOSP}_{max}$, it is equivalent 
to requiring all cycles to have strictly positive weight.
Then, any efficient set consists only of simple paths.

\begin{definition}[$\efflen$-Bounded]\label{def:EFFICIENCY_LENGTH}
    A \POSP{} instance $\instance$  is $\efflen$-bounded if there is some  $\efflen > 0 $ such that 
    %$|\p| \leq \efflen \, \forall \p \in \bigcup_{\vrt\in\V}\eff_{s,\vrt}$.
    $$
    \forall  \vrt \in \V \, \forall \p \in \PS_{s,\vrt} : \quad  |\p| > \efflen \implies  \exists \palt \in \PS_{s,\vrt} : |\palt| \leq \mu \land \palt \prec \p.
    $$
\end{definition}

\section{Optimal Substructure}\label{sec:dpp}

Labeling algorithms for \MOSP{} construct efficient (simple) paths in an iterative process.
This is possible since \MOSP{} exhibits \emph{optimal substructure}.
This optimal substructure is often identified as \emph{subpath optimality},
i.e., the property that every efficient path contains only efficient subpaths.
When considering \POSP{}, we find that there are several applications in which 
subpath optimality does not hold, but that can still be tackled by \MOSP{}-like labeling algorithms 
\citep{Euler2022, Dean2004, Gandibleux2006}.
On the other hand, \MOSP{} comes with certain assumptions baked into its weight structure that must be made 
explicit if one wants to identify \POSP{} instances solvable by labeling methods.
In the following, we 
develop various conditions on \POSP{} instances that
result in an optimal substructure on which labeling methods can be based.

\subsection{History-Freeness}

The \emph{history-freeness} property ensures that the expansions of two equivalent paths along the same arc remain equivalent.

\begin{definition}[History-Free]\label{def:histfree}
    An instance $\instance$ is called \emph{history-free} if  for all $v \in V$
    and $\p,\palt\in\PS_{s,v}$ 
    it holds that
    \begin{align}
    \WW(\p) = \WW(\palt) \implies \WW(\p + \arc)  = \WW(\palt + \arc) \quad \forall \arc \in \outgoing{v}.
    \end{align}
\end{definition}

In many applications, e.g., time-dependent shortest path problems, history-freeness holds naturally.
In  others, e.g., in price-optimal public transit routing \citep{Euler2022}, history-freeness is only achieved after defining a tailored weight space.
History-freeness of $\instance$ implies the existence of an \emph{arc update function} \mbox{$\UF: \VS \times \A \to \VS$}
such that for all paths $\p=(\vrt_1,\dots,\vrt_n) \in  \PS_s$ we have \mbox{$\WW(\p) = \UF(\WW((\vrt_1,\dots,\vrt_{n-1})), (\vrt_{n-1},\vrt_n) )$}, i.e., the update on
a fixed arc $a \in \A$ depends solely on the weight of the path and not on its node set.
It is often more handy to define the weight function $\WW$ via $|\A|$ update functions than by listing paths explicitly.
In the following, we will rely on this mode of presentation whenever it is convenient.

\subsection{Independence \& Subpath Optimality}

In the classical shortest path problem (\SP{}) and \MOSP{},
the correctness of labeling methods is usually argued via \emph{subpath optimality}.
In \SP{}, subpath optimality is a form of Bellman's principle of optimality \citep{Bertsekas2012}.

\begin{definition}[Subpath Optimality] 
A path $\p= (v_1,\dots,v_k)  \in\PS$ is subpath optimal if
every subpath $(v_1,\dots,v_i), i\in[k]$ of $\p$ is efficient.
We say $\instance$ is subpath optimal if all efficient paths in $\PS_{s}$ are subpath optimal.
\end{definition}

When studying variants of \MOSP{}, authors usually try to prove subpath optimality in their application, too.
\cite{Perny2005}, for example, compute maximal complete sets of efficient simple paths on acyclic digraphs with respect to a
quasi-transitive binary relation on  $2^\A$ that fulfills the \emph{independence} property.
They then show that independence implies subpath optimality on acyclic digraphs.
Their results remain limited to acyclic digraphs because their binary relation is defined on simple paths only and, as \cref{ex:SPO_NOT_SUFFICIENT} shows, subpath optimality does not hold on the space of simple paths in general digraphs.
We therefore extend the notion of independence of \cite{Perny2005} by considering a partial order on $\VS$ instead. 
This allows us to tackle $\POSPMIN{}$ and $\POSPMAX{}$ on general digraphs.
As noted in \cref{sec:posp}, this is equivalent to considering a preorder on $\PS.$

\begin{definition}[Independence {\citep[cf.][]{Perny2005}}]\label{def:independence}
An instance $\instance$ is called \emph{independent} if
for all $v \in \V$ and $\p,\palt \in \PS_{s,v}$ it holds that
\begin{align}
\WW(\p) \strictpo \WW(\palt) \Rightarrow  \WW(\p + \arc) \strictpo \WW(\palt + \arc) \quad \forall \arc \in \outgoing{v}.
\end{align}
\end{definition}

\begin{example}\label{ex:SPO_NOT_SUFFICIENT}
Consider the instance $\instance$ 
given by the graph in \cref{fig:SPO_SWO}{.a} and the partial order 
on the weight space $\VS=\{A,B,C\}$ given by the Hasse diagram in \cref{fig:SPO_SWO}{.b}.
The weight function is given by $\WW(st) = C$,
$\WW(s) =  \WW(s\vrt) = \WW(s\vrt t) = \WW(s\vrt t \vrt) = \dots = A $ and 
$\WW(s t \vrt ) = \WW(s t \vrt t ) = \dots = B $.
%
%Weights are given as follows:
%\begin{alignat*}{2}
%\WW(s) = A                    &&                    \\
%\WW(s\vrt) = A                && \quad \WW(st) = C  \\
%\WW(s\vrt t) = A              && \quad \WW(stv)=B   \\
%\WW(s\vrt t \vrt) = A         && \quad \WW(stvt)=B  \\
%%\WW(s\vrt t \vrt t ) = A      && \quad \WW(stvtv)=B \\
%        \dots                && \dots             \\ 
%\end{alignat*}
It is easy to see that $\instance$ is indeed well-posed, history-free, and independent.
\begin{figure}[ht]
\centering
\subfloat[Graph]{
    \parbox[b][2cm][t]{.49\textwidth}{
     \centering
	\begin{tikzpicture}[scale=1]
	\node[] (s) at (-2,0){\small$s$};	
	\node[] (v) at (0,0){\small$\vrt$};	
	\node[] (t) at (2,0){\small$t$};	
        \draw[garc] (s) to[bend right=45] (t) {};
	\draw[garc] (s) -- (v)  {};
        \draw[garc] (t) to [bend right=45] (v) {};
        \draw[garc] (v) to (t) {};
        \end{tikzpicture}
}}
%\end{minipage} 
%\begin{minipage}{.45\linewidth}	
%        \centering	
\subfloat[Weight Space]{
    \parbox[b][2cm][t]{.49\textwidth}{
    \centering
        \begin{tikzpicture}[scale=0.9]
        \node[] (A) at (-2,0){\small$A$};	
        \node[] (B) at (0,0){\small$B$};	
        \node[] (C) at (2,0){\small$C$};	
        \draw[warc] (B) to (A) ;
        \path[garc] (A) to[bend right=45] (C) {};
        \path[garc] (C) to [bend right=45] (B) {};
        \end{tikzpicture}}}
%\end{minipage}
%\medskip 
\caption{The graph and weight space used in \cref{ex:SPO_NOT_SUFFICIENT}.} \label{fig:SPO_SWO}
\end{figure}
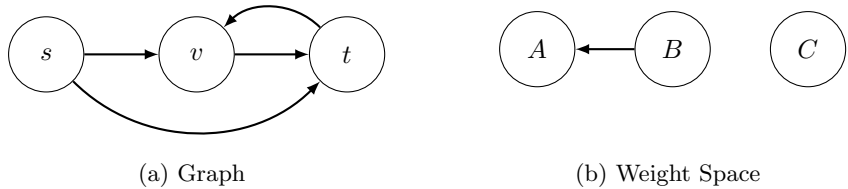
There are exactly two simple $s,t$-paths $(s \vrt t)$ and $(st)$. They are of weight $\WW((s\vrt t))= A$ and 
$\WW((st)) = C$, respectively; both are not dominated by any other simple path.
However, $(s \vrt t)$ contains the subpath $(s \vrt)$ that is dominated by $(s t \vrt)$.
Hence, simple paths that are efficient w.r.t. $\SPS_{s,t}$ may indeed contain non-efficient simple subpaths.
However, since $\WW((st\vrt t)) \strictpo \WW((s\vrt t))$, $(s\vrt t)$ is not efficient w.r.t. $\PS_{s,t}$
and thus every possible set $\effopt_{s,t}$ must contain a non-simple path.
\end{example}

Using \cref{def:independence}, independence does indeed imply subpath optimality in general digraphs. 
\begin{proposition}\label{thm:ISWO}
Any independent instance $\instance$ is subpath optimal.
%If an instance $\instance$ is independent, then it is also subpath optimal.
\end{proposition}
\begin{proof}
    Assume there is an efficient path $\p=(s,\dots,\vrt,\dots,t) \in\PS_{s,t}$ containing a dominated subpath $(s,\dots,\vrt)$.
    Then, there is another $s,\vrt$-path $\palt$ with
    $ \WW(\palt) \strictpo \WW((s,\dots,\vrt))$.
    By repeatedly applying independence,
    we obtain the contradiction $\WW(\palt + (\vrt,\dots,t)) \strictpo \WW(\p)$.
\end{proof}

\begin{rem}
\Cref{thm:ISWO} holds even if $\instance$ is not history-free.    
\end{rem}

For some \POSP{} instances common in practice \citep{Dean2004,Gandibleux2006,Euler2022, Smith2012,Strehler2017}, independence does not hold.
For example, minimizing the number of fare zones traversed by a public transit journey can be modeled by an instance 
$\instance$ with weights drawn from the $n$-element subset lattice, i.e.,
$(\VS,\po) = (2^{[n]},\subseteq)$,
$\WW((\vrt)) := \emptyset \, \forall \vrt\in\V$,
and  $\UF(\tau, \arc) := \tau \cup \mathbb{X}_{\arc} \, \forall \tau\in\VS,\arc\in\A$ 
for some $\mathbb{X}\in (2^{[n]})^\A$.

Then, given paths $\p,\palt \in \PS_{s,t}$ with $\WW(\p) = \{1,2\}$ and $\WW(\palt) = \{1,2,3\}$ and
$\arc\in\outgoing{t}$ with $\mathbb{X}_{\arc} = \{3\}$,
we have $\WW(\p)\subset \WW(\palt)$ but
\begin{equation}
\WW(\p + \arc) = \UF(\{1,2\},\arc) = \{1,2,3\} = \UF(\{1,2,3\},\arc) =   \WW(\palt + \arc).    
\end{equation}
Hence, the dominated path $\p$ has a chance to ``catch up''  and still be a subpath of an efficient path.
In this and similar cases, the following relaxed form of independence, \emph{weak independence}, applies.
Note that it is not related to the relaxation of the same name proposed by \cite{Perny2005}.
Weak independence does not imply subpath optimality.
Under some additional assumptions, however, a weaker form may apply.

\begin{definition}[Weak Independence]
    An instance $\instance$ is \emph{weakly independent} if 
    for all $v \in \V$ and $\p,\palt \in \PS_{s,v}$ it holds that
    \begin{align}
    \WW(\p) \strictpo \WW(\palt) \Rightarrow  \WW(\p + \arc) \po \WW(\palt + \arc) \quad \forall \arc \in \outgoing{v}. 
    \end{align}
\end{definition}

\begin{definition}[Weak Subpath Optimality]
An efficient path $\p  \in\PS_{s,t}$
is weakly subpath optimal if there exists an $s,t$-path $\palt=(s=v_1,\dots,v_k=t)$ with $\WW(\p) = \WW(\palt)$
such that every subpath $(v_1,\dots,v_i), i\in[k]$ of $\palt$ is efficient.
If all efficient paths in $\PS_s$ are weakly subpath optimal, $\instance$ is weakly subpath optimal.
\end{definition}

In \cref{section:WSOSimple,section:WSOCycle}, 
we will identify properties of weakly independent instances that
imply weak subpath optimality.
For all proofs, the core idea is the following construction argument, in which we build a sequence
of efficient paths coming increasingly closer to weak subpath optimality.

\subsection{A Constructive Argument}\label{construction:PROOF}

We consider a history-free and weakly independent  instance $\instance$.
Further, assume that every chain in $\image(\WW)$ has a minimum element.
Given an efficient path $\p\in\PS_{s,t}$ with $\p = (v_1,\dots,v_{n})$, let $k(\p) \in [n-1]_0$ be the largest number such that $(v_1,\dots,v_{n-l})$
is efficient for all $l \in[k(\p)]_{0}$.
If $k(\p) = n-1$, $\p$  is subpath optimal.

Now, we  construct a sequence of $s,t$-paths $(\p_i)_{i\in I}$  starting with any efficient $s,t$-path $\p_1\in\PS_s$.
Assuming we have already constructed the paths $\p_1,\dots,\p_{i}$, we continue the sequence as follows.
Let $\p_i=(v_1,\dots,v_n)$.
If $k(\p_i) < n-1$, then there must be a path $\palt=(u_1,\dots,u_m)$ with $u_1=v_1$ and $u_m=v_{n-k(\p_i)-1}$
such that $\WW(\palt) \strictpo \WW(v_1,\dots,v_{n-k(\p_i)-1})$.
Since every chain  in $\image(\WW)$ has a minimum element, we can choose $\palt$ such that there is no other $v_1,v_{n-k(\p_i)-1}$-path $\paalt$
with $\WW(\paalt) \strictpo \WW(\palt)$.
Weak independence now implies
%\begin{align}\label{eq:wso_proo}
$    \WW \left(\palt + \left(v_{n-k(\p_i)-1},v_{n-k(\p_i)}\right)\right) \po \WW\left( \left( v_1,\dots,v_{n-k(\p_i)} \right)\right).$
%\end{align}
Since $ \left( v_1,\dots,v_{n-k(\p_i)} \right)$ is efficient, we have
\begin{equation}
\WW(\palt + (v_{n-k(\p_i)-1},v_{n-k(\p_i)})) = \WW((v_1,\dots,v_{n-k(\p_i)})).    
\end{equation}
By history-freeness now all paths $\palt + (v_{n-k(\p_i)-1},\dots,v_{n-l}), l\in[k(\p_i)]_0$ must be efficient, too.
Hence, \mbox{$\p_{i+1} := \palt + (v_{n-k(\p_i)-1},\dots,v_{n})$} is efficient and 
$k(\p_{i+1}) > k(\p_i)$.
The sequence $(k(\p_i))_{i\in I}$ is strictly increasing
and $\WW(\p_i) = \WW(\p_j)$ for $i,j\in I$.
Thus, we either have $I=\N$ or $I = [m]$ for some $m\in\N$.
In the latter case, $\p_m$ is a subpath optimal $s,t$-path, and $\p_1$ is a  weakly subpath optimal $s,t$-path of the same weight.

\Cref{example:NEG_CYCLE} shows that a well-posed, history-free and weakly independent instance may not be 
weakly subpath optimal.
In that case, $I = \N$.

\begin{example}\label{example:NEG_CYCLE}
    Consider the instance $(\G,\WW,\VS,\po,v_1)$ with $\G=(\V,\A)$ given by
    $\V=\{ \vrt_1,\vrt_2\}$ and  $A = \{ (\vrt_1,\vrt_2), (\vrt_2,\vrt_2) \}$.
    The weight function $\WW$ is defined as
    $\WW(\vrt_1,\vrt_2) := 2$  
    and $\WW(\vrt_1,\vrt_2,\vrt_2) = \WW(\vrt_1,\vrt_2,\vrt_2,\vrt_2) = \dots =  1$
    with $\VS= \N$.
    Here, the only dominated $\vrt_1,\vrt_2$-path is $(\vrt_1,\vrt_2)$.
    All other paths have a weight of $1$ and are efficient. Hence, $\WW$ is well-posed, history-free and weakly independent.
    However, all efficient $\vrt_1,\vrt_2$-paths contain the dominated path $(\vrt_1,\vrt_2)$.
    Thus, with $\p_1 = (\vrt_1,\vrt_2,\vrt_2)$ the sequence $(\p_i)_{i\in I}$
    is of the form $\p_i = (\vrt_1,\vrt_2,\dots,\vrt_2)$ with $\vrt_2$ appearing $i+1$ times. Hence, $I=\N$.
\end{example}

\subsection{Weakly Subpath Optimal Simple Paths}\label{section:WSOSimple}

In this section, we analyze the conditions on \POSP{} that ensure that an efficient and \emph{simple} path exists.
To do so, we need the following lemma.

\begin{lemma}\label{lemma:WALK_TO_PATH}
    Consider an instance $\instance$ that is  history-free, weakly independent and cycle-increasing.
    Then, for all $t \in \V$ and every $s,t$-path $\p\in\PS_{s,t}$, there is a simple $s,t$-path $\palt \in \SPS_{s,t}$
    with $\WW(\palt) \po \WW(\p)$.
    Moreover, $\instance$ is well-posed.
\end{lemma}

\begin{proof}
    Consider the non-simple $s,t$-path $\p = (s,\dots,v,\dots,v,\dots,t)$.
    Since $\WW$ is cycle-increasing, we must have
    $ \WW(s,\dots,v) \po \WW(s,\dots,v,\dots,v)$. 
    Applying weak independence and history-freeness yields $\WW(s,\dots,v,\dots,t) \po \WW(\p)$.
    We set $\p \leftarrow (s,\dots,v,\dots,t)$ and repeat this process until $\p$ is a simple path.
    This proves the first claim.
    For the second claim, $|\bigcup_{\vrt\in \V}\nondom_{s,\vrt}| < \infty$ follows immediately from the first.
    %Short:
    %Further, any chain $\chain \subseteq \WW(\PS_s)$ must have a minimum element by cycle-increasingness of $\instance$.
    Any chain $\chain \subseteq \WW(\PS_s)$ without minimum element  is induced by an infinite sequence of paths
    $(p_i)_{i\in I}$ with \mbox{$\WW(\p_i) \strictsuccpo \WW(\p_{i+1})$}.
    This sequence must contain at least one infinite subsequence of s,t-paths for some $t\in \V$ with strictly increasing length.
    As this violates cycle-increasingness, all chains must have a minimum element.\end{proof}

\begin{proposition} \label{thm:WIPSO}
    An instance $\instance$ that is history-free, weak\-ly independent and cycle-increasing is weakly subpath optimal.
    Moreover, for all $t\in \V$ and every  $\va \in \nondom_{s,t}$, there is a simple path $\p\in\SPS_{s,t}$ with
    $\WW(\p) = \va$.
\end{proposition}

\begin{proof}
Let $\p=(s=v_1,\dots,v_n=t)\in \PS_{s,t}$ be an efficient $s,t$-path.
By  \cref{lemma:WALK_TO_PATH}, $\p$ is w.l.o.g. a simple path.
Since every chain has a minimum by  \cref{lemma:WALK_TO_PATH}, we can follow  \cref{construction:PROOF}
with the following modification:
We always choose $\palt \in \argmin \{ \WW(\p) \vert \, \p \in \PS_{v_1,v_{n-k(\p)-1}}\}$ as a simple path.
This is again possible by \cref{lemma:WALK_TO_PATH}.
Now, we claim that $(\p_i)_{i\in I}$ is a sequence of simple paths.
Since $k((\p_i)_{i\in I})$ is strictly increasing, $I$ must then be finite.
With $\WW(\p_i) = \WW(\p)$ for all $i\in I$, this concludes the proof.

We argue the claim via induction over $I$.
For some $i\in I$, assume  \mbox{$\p_i = (v_1,\dots,v_n)$} is a simple path, but 
$\p_{i+1} = \palt + (v_{n-k(\p_i)-1}, \dots ,v_n)$ with $\palt= (u_1 = v_1,\dots,u_m = v_{n-k(\p_i)-1})$
is not.
Since the subpaths $\palt$ and $(v_{n-k(\p_i)-1}, v_{n-k(\p_i)},\dots, v_n)$ are simple,
there must be some $r\in\{2,\dots,m-1\}$ and $l\in\{n-k(\p_i),\dots,n\}$ such that $u_r=v_l$.
Remember that by definition of $\palt$
\begin{equation}  \label{eq:PSO_PROOF_EQ_1}
 \WW(u_1,\dots,u_r,\dots,u_m) \strictpo \WW(v_1,\dots,v_{n-k(\p_i)-1}).
\end{equation}

By weak independence, it holds that
\begin{equation}  \label{eq:PSO_PROOF_EQ_2}
 \WW(u_1,\dots,u_r,\dots,u_m=v_{n-k(\p_i)-1},v_{n-k(\p_i)},\dots,v_l) \po \WW(v_1,\dots,v_l).   
\end{equation}
Now, the path $(u_1,\dots,u_r,\dots,u_m=v_{n-k(\p_i)-1},v_{n-k(\p_i)},\dots,v_l)$ contains the cycle
$(u_r,\dots,u_m=v_{n-k(\p_i)-1},v_{n-k(\p_i)},\dots,v_l=u_r)$
and by $\instance$ cycle-increasing, we have
%\begin{equation} \label{eq:PSO_PROOF_EQ_3}
$\WW(u_1,\dots u_r) \po \WW(v_1,\dots,v_l).$
%\end{equation}
By $(v_1,\dots,v_l)$ being efficient, it must be that
$\WW(u_1,\dots u_r) = \WW(v_1,\dots,v_l)$.
Thus, we append both paths with $(u_r,\dots,u_m)$ and obtain
\begin{multline} \label{eq:PSO_PROOF_EQ_4}
% \begin{aligned}
    \WW(u_1,\dots u_r,\dots,u_m) = \\ \WW(v_1,\dots, v_{n-k(\p_i)-1}  , \dots, v_l=u_r,u_{r+1},\dots,u_m = v_{n-k(\p_i)-1}) \\
                                 \succpo \WW(v_1,\dots, v_{n-k(\p_i)-1}  )
%\end{aligned}   
\end{multline} 
where the equality follows by history-freeness and the inequality by cycle-increasingness of $\WW$.
 \Cref {eq:PSO_PROOF_EQ_1,eq:PSO_PROOF_EQ_4} form a contradiction.
%\begin{figure}[ht]
%\centering
%\begin{tikzpicture}[scale=0.9]
%	\node[minimum size=1.9cm] (s) at (0,0){\tiny$v_1 = u_1$};	
%        \node[minimum size=1.9cm] (v) at (3,0){\tiny
%        $\begin{aligned}
%        v_{n-k(\p_i)-1} \\= u_m        
%        \end{aligned}$
%        };	
%	\node[minimum size=1.9cm] (u) at (6,0){\tiny$v_s=u_r$};	
%	\node[minimum size=1.9cm] (t) at (9,0){\tiny$v_n$};	
%        \draw[garc] (s) to[] (v) {};
%	\draw[garc] (v) -- (u)  {};
%        \draw[garc] (u) to (t) {};
%        \draw[garc, dashed] (s) to[ bend right = 30 ] (u) {};
%        \draw[garc,dashed] (u) to[ bend right = 30 ] (v) {};
%\end{tikzpicture}
%\caption{The digraph used in the proof of \cref{thm:WIPSO}. 
%The dotted arcs represent the path $(u_1,\dots,u_r)$; the solid arcs represent the path $(\vrt_1,\dots,\vrt_n)$.
%\label{fig:PSWO_PROOF}}
%\end{figure}
\end{proof}

\subsection{AHW instances}

We call arc-increasing, history-free and weakly independent \POSP{} instances AHW.
They are always well-posed and weakly subpath optimal.
AHW instances are structurally close to \MOSP{} instances in that an efficient complete set
can be generated from simple paths only. We will see in \cref{sec:algorithms} that they can be solved by both label-setting and label-correcting methods.
Note that the converse is not true, however, and instances lacking any of the three properties may still be solvable with
label-setting or label-correcting methods.
In the following, we analyze AHW instances that have been studied in the literature.

\subsubsection{Time-Dependent Shortest Path}\label{subsec:motdsp}

In the time-dependent shortest path problem \citep{Cooke1966,Orda1990, Dean2004, Foschini2012}, we aim to find a shortest simple path in a directed graph $\G=(\V,\A)$ given a starting station $s\in V$, a starting time $\tau_0\in \R_{\geq 0}$ and a
travel-time function \mbox{$t: \A \times \R_{\geq 0} \to \R_{\geq 0}$}.
The  function $t$ is assumed to have the 
\emph{First-In-First-Out}-property (\emph{FIFO}), i.e.,
\begin{equation}
  \forall \arc\in\A \, \forall \tau_1,\tau_2\in\R_{\geq 0}  \quad \tau_1 \leq \tau_2 \implies \tau_1 + t(a,\tau_1) \leq \tau_2 + t(a,\tau_2).  
\end{equation}
The weight function $\WW$ on $\PS_s$ is given by $\WW((s)) := \tau_0$ 
and the arc update function
\mbox{$\UF: \A \times \R_{\geq 0} \to \R_{\geq 0}$} 
$(a,\tau) \mapsto \tau + t(a,\tau)$.
The resulting \POSP{} instance $(\G,\WW,\R_{\geq 0},\leq,s)$ is weakly independent by 
the FIFO property of $t$ and arc-increasing by $t$ mapping into $\R_{\geq 0}$.
Hence, time-dependent shortest path problems are AHW.

\subsubsection{Partially Ordered Positive Monoids}\label{subsec:pomonoid}

\cite{Euler2022} study a \POSP{} application in public transportation where the price of a (simple)
path depends on multiple criteria, like fare, zones that cannot be represented in a \MOSP{} framework.
The authors propose \emph{partially ordered, positive monoids} as an appropriate weight structure.

Consider a \POSP{} instance $\instance$ for which
there is a binary operator $+$ such that $(\VS,+,\po)$ is a posemigroup and 
$\UF: \: \VS\times\A \to \VS \:\: (\va,\arc) \mapsto \va + \mathbb{X}_{\arc}$
for some $\mathbb{X}^\A \in \VS^\A$.
$(\VS,+,\po)$ is a posemigroup if $+$ is associative and translation-invariant w.r.t to $\po$, i.e.,
\begin{equation}\label{eq:transin}
\sigma \po \va \Rightarrow \sigma+\upsilon \po \va+\upsilon \, \forall \sigma,\va,\upsilon \in \VS.
\end{equation}
\Cref{eq:transin} and the definition of $\UF$ directly imply history-freeness and weak independence.
If there is additionally an identity element $e\in\VS$, $(\VS,+,\po)$ is a partially ordered monoid.
It is called \emph{positive} if \mbox{$e\po \va \, \forall \va \in \VS$}.
We then have for all $t\in\V$ and  $\p\in \PS_{s,t}$,
\begin{equation}
    e \po  \mathbb{X}_{\arc}  \implies e + \WW(\p) \po \WW(\p) + \mathbb{X}_{\arc}
    \implies \WW(\p) \po  \WW(\p) + \mathbb{X}_{\arc}
                   \quad \forall \arc \in \outgoing{t}.
\end{equation}
Hence, $\instance$ is AHW.

\subsubsection{Multi-Objective Bottleneck Weight Functions}\label{subsec:bottleneck}
% join-semilattices
% function diagonals

Consider the multi-objective bottleneck problem studied by \cite{Gandibleux2006}, i.e.,
a \POSP{} instance $(\G,\WW,\R^{[m+n]}_{\geq0},\trianglelefteq,s )$ with $\WW=\WW_1\Delta\WW_2$ such that
there is some $(\mathbb{X}^1,\mathbb{X}^2) \in \R_{\geq0}^{[m]\times \A} \times \R_{\geq0}^{[n]\times \A}$
and
\begin{equation}
\UF( (\va_1,\va_2) ,\arc) = 
(\UF[\WW_1](\va_1,\arc), \UF[\WW_2](\va_1,\arc)) = 
(\va_1 +  \mathbb{X}^1_{\arc},  \min(\va_2,\mathbb{X}^2_{\arc})).
\end{equation}
Here,  the $\min$ operation is understood as pairwise.
The partial order operator $\trianglelefteq$ is defined as
%\begin{equation}
$(\va_1,\va_2) \trianglelefteq (\vao_1,\vao_2) \iff \va_1 \leq \vao_1 \land  \va_2 \geq \vao_2.$ 
%\end{equation}
First, note \mbox{$(\G,\WW_1,\R^{[m]}_{\geq0},\leq,s )$} is a \MOSP{} instance and therefore AHW.
To see that $(\G,\WW,\R^{\{m+1,\dots,m+n\}}_{\geq0},\geq,s )$ is AHW, we first show that weight structures derived from join-semilattices result in AHW'ness.

A join-semilattice $(\VS, \join)$ consists of a ground set $\VS$ and 
an associative, commutative and idempotent binary operator $\join$ on $\VS$.
On $(\VS, \join)$, a canonical partial order is given by 
$\va \po \valt :\Longleftrightarrow \valt = \va \join \valt \, \forall \va,\vao \in\VS$.
Following \cite{Davey2002},
we have 
%\begin{equation}
$\va \po \valt \Rightarrow  \va \join \vao \po \valt \join \vao \, \forall \va,\valt,\vao \in \VS.$ 
%\end{equation}
Hence,  $(\VS, \join,\po)$ is a posemigroup.
Any instance $\instance$ with $\WW$ given by
%\begin{equation}
$\UF: \: \VS\times\A \to \VS \:\: (\va,\arc) \mapsto \va \join \mathbb{X}_{\arc}$
%\end{equation}
for some $\mathbb{X}\in\VS^\A$ is thus history-free and weakly independent, as shown in \cref{subsec:pomonoid}.
By $\va \po \va \join \vao \, \forall \va,\vao\in\VS$,
it is also arc-increasing and, therefore, AHW.

Now, note that $(\R_{\geq0}^m, \min)$ is a join-semilattice as $\min$ is an associative, commutative and idempotent binary operator;
$\geq$ is the canonical partial order on $(\R_{\geq0}^m, \min)$.
Consequently, $(\G,\WW_2,\R^{\{m+1,\dots,m+n\}}_{\geq0},\geq,s)$ is AHW.

Finally, it is easy to see that any instance  $(\G, \WW_1 \Delta \WW_2, \VS_1 \times \VS_2, \po,s)$
with  $ (\va_1,\va_2) \po (\vao_1,\vao_2) \iff  \va_1 \po_1 \vao_1 \land \va_2 \po_2 \vao_2 $
is AHW if and only if $(\G,\WW_1,\VS_1,\po_1,s)$ and $(\G,\WW_2,\VS_2,\po_2,s)$ are both AHW.
Hence, the \POSP{} instance $(\G,\WW,\R^{[m+n]}_{\geq0},\leq,s )$ is AHW.

\subsubsection{Interval Weights}\label{subsec:interval}
\cite{Okada1994} study a multi-objective shortest path problem 
where arc weights are closed intervals on $\R_{\geq 0}$.
An interval $\va \in \VS$ is represented by the tuple $(\va_c,\va_w)$ of its center and width.
Since we only consider intervals in the nonnegative reals, it holds $\va_c / \va_w \geq 1$.
The addition on $\VS$ is given by 
$\va+\valt := (\va_c+\valt_c,\va_w+\valt_w)$ and for fixed $-1\leq \alpha \leq \beta \leq 1$ a partial order operator $\po_{\alpha,\beta}$ on $\VS$ is given by
\begin{equation}
 \begin{aligned}
    \va \po_{\alpha,\beta} \valt \iff &   \alpha(\va_w-\valt_w)   \leq  \valt_c - \va_c  \land
                                          \beta(\va_w-\valt_w)   \leq  \valt_c - \va_c.
\end{aligned}   
\end{equation}
This problem can then be formalized as a \POSP{} instance
$(\G,\WW,\VS,\po_{\alpha,\beta},s)$
with $\WW((s)) := (0,0)$
and  
%\begin{equation}
$\UF: \: \VS\times\A \to \VS \:\: (\va,\arc) \mapsto \va + \mathbb{X}_{\arc}$ 
%\end{equation}
for some $\mathbb{X}\in \VS^\A$.
The instance is clearly history-free. 
Independence and arc-increasingness can be shown by basic calculations (see \ref{appendix:okada}).
Hence,   $(\G,\WW,\VS,\po_{\alpha,\beta},s)$ is not only AHW but also subpath optimal.

\subsection{Weakly Subpath Optimal Paths with Cycles}\label{section:WSOCycle}

In AHW instances, a complete set of efficient paths can always be constructed from simple paths only.
In the following, we study weak subpath optimality for instances where 
paths that contain cycles are essential for obtaining such a set. 
These are necessarily not arc-increasing.

\begin{proposition} \label{thm:WIPSO2}
An instance  $\instance$ that is history-free, weak\-ly independent
and $\efflen$-bounded is also weakly subpath optimal.
%Then, $\instance$ is weakly subpath optimal.
\end{proposition}

\begin{proof}
    Let $\p\in \PS_{s,t}$ be an efficient $s,t$-path. 
    We want to show that there is a subpath optimal
    path $\palt\in\PS_{s,t}$ with $\WW(\p) = \WW(\palt)$.
    Starting with $\p_1 = \p$, we employ  \cref{construction:PROOF} 
    to obtain a sequence of efficient $s,t$-paths $(\p_i)_{i\in I}$ with $\WW(\p_i) = \WW(\p) \, \forall i \in I$.
    This is possible since $\efflen$-boundedness directly implies that every chain
    in $\image(\WW)$ has a minimum element.
    By construction, we also have that $k(\p_i)_{i\in I}$ is strictly increasing.
    Hence, no path can appear twice in  $(\p_i)_{i\in I}$.
    Additionally, $\efflen$-boundedness implies that the set of all efficient paths is finite.
    Thus, $I$ must be a finite set, say $I = [j]$. Then, $\p_j$ must be subpath optimal.
\end{proof}

\begin{proposition}\label{thm:WIPSO3}
    Consider an instance $\instance$ that is history-free, weak\-ly independent and cycle-non-decreasing.
    If $\instance$ is also well-posed, it is weakly subpath optimal.
\end{proposition}

\begin{proof}
    Let $\p\in \PS_{s,t}$ be an efficient $s,t$-path. 
    To show that there is a subpath optimal path $\p_j\in\PS_{s,t}$ with $\WW(\p) = \WW(\p_j)$,
    we employ \cref{construction:PROOF} beginning with $\p_1 := \p$
    to obtain a sequence of efficient $s,t$-paths $(\p_i)_{i\in I}$ with $\WW(\p_i) = \WW(\p) \, \forall i \in I$.
    The next path $\p_{i+1}$ can always be chosen, since we explicitly assume all chains to have a minimal element.
    By \cref{construction:PROOF}, $k(\p_i)_{i\in I}$ is strictly increasing, and 
    no path can appear twice in  $(\p_i)_{i\in I}$.
    It remains to show that $I$ is indeed a finite set. 

    To do so, assume $I$ is infinite.
    Then, there must be some $v\in\V$ such that the substitute path $\palt$ is chosen as an $s,v$-path an infinite number of times.
    Since  \cref{construction:PROOF}  always selects an efficient substitute path and $|\nondom_{s,\vrt}|<\infty$, there must be one path
    $\palt \in \PS_{s,t}$ that is selected an infinite number of times.

    By aggregating possible intermediate iterations, 
    we can assume w.l.o.g. that $\palt$ is selected in iterations $i$ and $i+2$.
    Then, we write $\p_i = (v_0,\dots,v_n)$ and $\palt = (v_0,\dots,v_h)$ with $h < n$.
    In iteration $i+1$, we substitute some subpath $(v_0,\dots,v_l)$ of $\p_i$
    with $\paalt = (u_0,\dots,u_m)$ where $v_0=u_0$, $v_l = u_m$ and $l < h$.
    Since, we reintroduce $\palt$ again in iteration $i+2$, there must be some $r \in \{0,\dots,m-1\}$ such that
    $u_r = v_h$.
    Now, we have that

    \begin{equation}
    \begin{aligned}
                    & \quad &&\WW(\palt) \strictpo \WW((u_0,\dots,u_r))                     &&\text{ since } \palt \text{ replaces } (u_0,\dots,u_r)\\
        \Rightarrow &       &&\WW(\palt + (u_r,\dots,u_m)) \po \WW(\paalt)                  &&\text{ by weak independence} \\
        \Rightarrow &       &&\WW(\palt + (u_r,\dots,u_m)) = \WW(\paalt)                    &&\text{ since $\paalt$ efficient } \\
        \Rightarrow &       &&\WW(\palt + (u_r,\dots,u_m)) \strictpo \WW((v_0,\dots,v_l))   &&\text{ since $\WW(\paalt) \strictpo \WW((v_0,\dots,v_l))$}.
    \end{aligned}
    \end{equation} 
    However,  this forms a contradiction to $\instance$ being cycle-non-decreasing as 
    $\palt + (u_r,\dots,u_m) = (v_0,\dots,v_l) + (v_l,\dots,v_h) + (u_r,\dots,u_m=v_l)$.
    Thus, $I$ must be a finite set, say $I = [j]$, and $\p_j$ subpath optimal.
\end{proof}

%\begin{rem}
%    While for \cref{thm:WIPSO,thm:WIPSO2}, well-posedness of $\instance$ follows from  \cref{lemma:WALK_TO_PATH} and $\efflen$-boundedness, respectively,
%    it has to be specifically required for  \cref{thm:WIPSO3}.
%\end{rem}

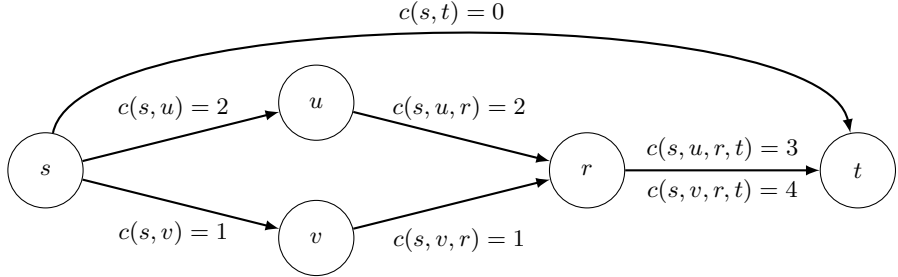
\begin{figure}[t]
\footnotesize
\centering
\begin{tikzpicture}[scale=0.9]
    \node[] (s) at (0,0){$s$};	
    \node[] (v) at (4,-1){$\vrt$};	
    \node[] (u) at (4,1){$u$};	
    \node[] (r) at (8,0){$r$};	
    \node[] (t) at (12,0){$t$};	
    \draw[garc] (s) -- node[labelnode,below,yshift=-.1cm,xshift=-.1cm] {$\WW(s,v) = 1$} (v);
    \draw[garc] (s) -- node[labelnode,above,yshift=.1cm,xshift=-.1cm] {$\WW(s,u) = 2$} (u);
    \draw[garc] (v) -- node[labelnode,below,yshift=-.2cm,xshift=.1cm] {$\WW(s,v,r) = 1$} (r);
    \draw[garc] (u) -- node[labelnode,above,yshift=.1cm,xshift=.1cm] {$\WW(s,u,r) = 2$} (r);
    \draw[garc] (r) -- node[labelnode,above,yshift=0cm] {$\WW(s,u,r,t) = 3$} node[labelnode, below, yshift=0cm] {$\WW(s,v,r,t) = 4$} (t);
    \draw[garc] (s) .. controls (0.5,2.5) and (11.5,2.5) .. node[labelnode,above,yshift=0cm] {$\WW(s,t) = 0$} (t); 
\end{tikzpicture}
\caption{
Consider the instance $(\G,\WW,\N,\leq,s)$ given by the graph above, in which
each path $\p$ starting in $s$ is annotated with a weight $\WW(\p)$ drawn from $\N$. 
The partial order $\leq$ is the natural total order on $\N$.
It is easy to see that $(\G,\WW,\N,\leq,s)$ is well-posed, history-free, weakly subpath optimal, and
even arc-increasing but not weakly independent.
}
\label{fig:WIPSO_DOES_NOT_IMPLY_WI}
\end{figure}  
Despite the previous results,
it is not true that weak subpath optimality is a stronger property than weak independence.
This is illustrated in \cref{fig:WIPSO_DOES_NOT_IMPLY_WI}.
We conclude this section with applications of \cref{thm:WIPSO2,thm:WIPSO3}.

\begin{example}[Electric Vehicle Shortest Paths with Recuperation and Recharging (EVSP)]\label{example:ELECTRIC}
    Due to their limited range, shortest path search for electric vehicles must take battery recuperation and recharging into account \citep{Baum2019}.
    Here, \emph{recuperation} refers to the recharging of the battery while traveling on downhill slopes, while
    \emph{recharging} means charging the battery at a (stationary) charging station or exchanging the whole battery.
    Battery swapping and recharging can result in efficient paths with cycles, as detours to charging stations might be necessary \citep{Strehler2017}.
    Several models for the EVSP have been discussed in the literature.
    If batteries can only be swapped and not recharged, simple paths between all battery swapping stations can be precomputed \citep{Storandt2012}.
    \citet{Strehler2017} discretize battery states in a battery-extended networking.
    \citet{Baum2019} propagate piece-wise linear charging functions through the graph. This results in a continuous Pareto front of charging and arrival time.
    We propose a simple (biobjective) \POSP{} model with discretized charging times that allows us to use realistic non-linear charging functions \citep[][]{Loebel2023}.
    
    We are given a graph $\G=(\V,\A)$ including some charging stations $C \subset \V$, a starting station $s\in\V$ and the initial \emph{state-of-charge} (SoC) 
    of the electric vehicle $\beta$ with $0 < \beta \leq 1$.
    Each arc $\arc\in\A$ has a time cost $t(\arc)\in\R_{>0}$ and a possibly non-linear charge function $ f_\arc: [0,1] \to [0,1]$ that models the
    change in the battery's SoC. 
    For any charging station $\vrt\in C$, we are given a non-linear, monotonously non-decreasing \emph{charge curve} $\xi_{\vrt}:[0, \infty) \to [0,1]$ mapping
    a charge time to the final SoC when charging an empty battery. From $\xi_{\vrt}$, we derive
    a charge increment function $\Delta \xi_{\vrt}(y,t) = \xi_{\vrt}(\xi_{\vrt}^{-1}(y)+t) -y$  \citep{Loebel2023}.
    At any charging station $\vrt\in \V$, the event of charging is modeled with a 
    loop arc $(v,v)$ corresponding to charging at $v$ for a discrete time step of $\epsilon > 0$ , i.e.,
    $t(v,v) = \epsilon$ and $f_{\vrt,\vrt}(y) = \Delta \xi_{\vrt} (y,\epsilon)$.
    All other arcs $\arc \in \A\backslash(C\times C)$ represent road segments with $f_\arc: y \mapsto \min(1,\max(0,y - y_\arc))$ where $y_\arc\in\R$ can be negative due to recuperation \citep{Baum2019}.
    An   EVSP instance $(G,c,\VS,\po,s)$ is then given by $\VS:= \R_{\geq0} \times [0,1]$,
    $\WW(s) :=(0,\beta)$, and $\po$ being the canonical partial order on $\VS \subseteq \R^2$. A path's cost is derived by  
    the arc update function $\UF: \VS \times \A\to \VS $ with
    \begin{equation}
        \UF((t,y), \arc) :=
        \begin{cases} 
        (t + t(\arc), f_\arc(y)) & \text{ if } y > 0\\
        (t + t(\arc), 0) & \text{ otherwise}.
        \end{cases}
    \end{equation}
    History-freeness of $\instance$ follows from defining $\WW$ via $\UF$.
    It is easy to see that weak independence holds. 
    Since every arc has positive travel time and the battery has a maximum capacity, the instance is also $\mu$-bounded. By \cref{thm:WIPSO2}, it is hence weakly subpath optimal.
    A minimal efficient set for EVSP might contain at most one infeasible path with an SoC of $0$.
    Such a path can never again become feasible and may hence be safely dropped by any labeling algorithm \citep[][]{Baum2019}.
\end{example}

\begin{example}[Weight-constrained shortest path with replenishment]\label{example:replenishment}
    \citet{Smith2012} introduced the \emph{weight-constrained shortest path problem with replenishment arcs} (WCSP-R).
    In it, we are given a directed graph $G=(V,A)$ in which every arc $a\in A$ carries a cost $w(a)\in \R_{\geq 0}$ and a weight $r(a)\in \R_{\geq 0}$.
    A subset of the arcs $\A'\subseteq A$ are \emph{replenishment arcs} that, when traversed, reset the weight to zero.
    The task is to find a minimum cost $s,t$-path of weight $< M$ for some $s,t\in \V$ and $M\in\R_{> 0}$.
    \citet{Smith2012} already observed that efficient $s,t$-paths may contain cycles.
    A common technique for solving resource-constrained path problems is to solve
    a biobjective version \citep[cf.][]{Ahmadi2022}.
    We can cast WCSP-R as a (biobjective) \POSP{} instance $\instance$ by letting
    $\VS := \R_{\geq 0} \times [0,M]$, setting initial costs 
    $\WW(s) := (0,0)$ and defining the update function for $(\va_1,\va_2)\in \VS$ and $\arc\in\A$ as

    \begin{equation}
        \UF((\va_1,\va_2), \arc) := \begin{cases}
            ( \va_1 + w(\arc) , M) & \text{ if } \va_2 + r(\arc)  \geq M\\
            ( \va_1 + w(\arc) , r(\arc))   & \text{ if } \va_2 + r(\arc) \leq M \text{ and } \arc \in \A'\\
            ( \va_1 + w(\arc) , \va_2 + r(\arc)) & \text{ otherwise.} \\
        \end{cases}
    \end{equation}
    The partial order $\po$ is the canonical order on $\VS\subset \R^2$.  
    The instance \mbox{$\instance$} is obviously history-free; weak independence is easy to show.
    It is, however, not cycle-non-decreasing: 
    There might be a cycle of zero cost that contains a replenishment arc.
    Traversing such a cycle twice, however, neither yields an improvement in cost nor weight.
    Hence, $|\bigcup_{\vrt\in\V} \nondom_{s,\vrt} | < \infty$.
    Now, assume there is a chain in $\image(\WW)$ without a minimum element.
    It must contain at least one infinite subsequence of $s,t$-paths of strictly increasing length for some $t\in\V$.
    This immediately leads to a contradiction, ensuring that $\instance$ is well-posed.
    If additionally every replenishment arc $\arc\in \A'$ has strictly positive costs $w(\arc)>0$, $\instance$ is cycle-non-decreasing.
    By \cref{thm:WIPSO3}, it is hence weakly subpath optimal.
    In case $w(\arc) > 0$ holds for all $\arc\in\A$, the instance is $\mu$-bounded,
    and weak subpath optimality can be derived from \cref{thm:WIPSO2}.
    Similar to \cref{example:ELECTRIC}, efficient sets contain at most one infeasible path of weight $M$.
   % As a weight of $M$ can no longer be replenished, it is safe to drop these paths in any labeling method \citep[cf.][]{Smith2012}.
\end{example}

\begin{example}[A $\max$-Objective Variation of Tourist Trip Planning]\label{example:TOURIST}
In the tourist trip planning problem \citep{Souffriau2010}, a tourist aims to find the most attractive route under a 
restriction on the total length of the route.
Tourist trip planning is frequently modeled as an orienteering problem \citep{Golden1987,Souffriau2010}.
Here, we are given a budget $B$ and a graph $\G=(\V,\A)$ in which each vertex $\vrt\in\V$ has a value $\rho_\vrt \geq 0$ and each arc $\arc\in\A$ a length $l_\arc > 0$.
The task is to find a path $(\vrt_1,\dots,\vrt_n) \in \PS_{s,t}$ maximizing
$\sum_{i\in [k]} \rho_{\vrt_i}$ under the restriction $\sum_{i\in[k-1]} l_{\vrt_{i},\vrt_{i+1}} \leq B$.
Note that an optimal path can contain loops.
Modeled as a (biobjective) \POSP{} instance, the orienteering problem is weakly independent
and cycle-non-decreasing, but not history-free. It is, indeed, not weakly subpath optimal.

To overcome this, we propose a slightly different model of the tourist trip planning problem.
We partition the vertices into $q$ disjoint categories of attractions $V = \bigcup_{i\in[q]}S_k$, e.g., restaurants,
historical sites, etc.
and aim to visit the most attractive sights in each category while respecting the length bound $B$.
The corresponding \POSP{} instance 
$(\G,\WW,\R^{q+1}_{\geq 0},\po,s)$
is given by letting 
\begin{equation}
c(\p) {:=} (\sum_{i\in[k-1]} l_{\vrt_{i},\vrt_{i+1}},\max_{ i\in[k]: \vrt_i\in S_1 } \rho_{\vrt_i}, \dots, \max_{ i\in[k]: \vrt_i\in S_q } \rho_{\vrt_i}) \end{equation}
 for every path $\p=(\vrt_1,\dots,\vrt_k)\in\PS$
and
\begin{equation}
(\va_1,\va_2) \po (\valt_1,\valt_2) \iff \va_1 \leq \valt_1 \land \va_2 \geq \valt_2
\end{equation}
for $(\va_1,\va_2),(\valt_1,\valt_2) \in \R_{\geq 0} \times \R^{q}_{\geq 0}$.
It is easy to verify that \mbox{$(\G,\WW,\R^{q+1}_{\geq 0},\po,s)$} is history-free and  weakly independent.
Note that it is also $\mu$-bounded since all cycles have positive length and after visiting all vertices no further improvement in any objective can be made.
Hence, by \cref{thm:WIPSO2} weak subpath optimality holds.
\end{example}

\section{Labeling Algorithms for \POSP}\label{sec:algorithms}

In the following, we provide a label-correcting and a label-setting algorithm for \POSP{}
based on the conditions developed in the previous section.
Depending on the \POSP{} variant and labeling technique, a different subset of them is required to ensure correctness.
For $\POSPMIN{}$, for example, 
the label-setting algorithm requires an instance to be weakly subpath optimal
while the label-correcting algorithm requires weak subpath optimality or weak independence but not necessarily both.
For $\POSPMAX{}$, history-freeness can be dropped for both algorithms, but subpath optimality must hold instead of weak subpath optimality.  The properties under which each algorithm can be applied for the different variants of \POSP{}
are summarized at the end of this section in \cref{tab:summary}.
Both types of algorithm maintain a tentative set of efficient $s,\vrt$-paths $\algeff_\vrt$ for every $\vrt\in\V$
that is iteratively updated with new paths.
In the label-setting case, paths can only be appended; in the label-correcting case, they can also be deleted if they are found to be dominated. 

In an implementation, the representation of the sets $\algeff_\vrt$ depends on the weight structure.
If a \POSP{} instance is history-free,
an $s,v$-path $\p$ is represented by a label $\la = (\la.vert,\la.pred, \la.val)$
consisting of the end vertex $\la.vert := \vrt$,
a pointer to the predecessor label $\la.pred$ and the weight $\la.val = \WW(\p)$.
The weight $\la.val$ is calculated recursively via 
an arc update function $\UF$, i.e., 
$\la.val =  \UF(\la.pred.val, (\la.pred.vert, \la.vert)).$
If $\instance$ is not history-free, however,  
no such function exists, and a full representation of the path must be stored.

The encoding length  of a $d$-dimensional \MOSP{} instance $(\G,\WW,\objSpace,\leq,s)$ is usually considered to be $\Theta(d|A|)$\citep[cf.][]{MaristanydelasCasasSedenoNodaBorndoerfer2021}.
Here, the structure of the update function $\UF$ and the weight space $(\objSpace,\leq)$ are not part of the encoding length.
For a  \POSP{} instance $\instance$, we also do not consider the encoding length of the weight function $\WW$ or the weight space $(\VS,\po)$ explicitly so that results on time complexity remain comparable to those for \MOSP{}.
Instead, we set the encoding length of the \POSP{} instance to $\Theta(\ENC |\A| )$ where $\ENC$ is a problem dependent parameter 
representing the weight structure. 
For $d$-dimensional \MOSP{} instances, we naturally let $\ENC:=d$.
Further, we assume that $\WW(\cdot)$ and $\po$ can both be evaluated in $\Theta(\ENC)$ time.
Under these assumptions, \POSP{} is, as a generalization of \MOSP{}, clearly intractable.

An \emph{output-sensitive running time bound} for an algorithm is a running time bound that depends on the input's and the output's encoding length. If this bound turns out to be polynomial w.r.t.\ to both encoding lengths, we call the algorithm \emph{output-sensitive} \citep{Boekler2018}.
As there are \emph{output-sensitive} algorithms for \MOSP{} \citep{Boekler2018, MaristanydelasCasasSedenoNodaBorndoerfer2021},
we check our algorithms for output-sensitivity, too.

\subsection{A Label-Correcting Algorithm}\label{sec:labelcorrecting}

With \cref{algo:corleymoon}, we introduce a label-correcting algorithm for \POSP{} based on the method of Bellman \& Ford \citep{Bellman1958}.
It is similar to a variant for \MOSP{}\, proposed by \cite{Corley1985} (see also \citep{Ehrgott2005}), but differs in the stopping criterion.
The algorithm maintains sets of tentative efficient paths in a set $\algeff_\vrt$ at each vertex $\vrt\in\V$.
After initialization, only the set $\algeff_s$ is populated.
In iteration $k$, a $\mathit{merge}$-operation then selects an efficient subset of $s,\vrt$-paths from
$\algeff_{\vrt,k-1} \cup \bigcup_{\vrtalt\in \ingoing{\vrt}} \{ \p + (\vrtalt,\vrt) | \p \in \algeff_{\vrtalt,k-1}  \}$
and assigns it to $\algeff_{\vrt,k}$.
%in \crefrange{line:cm_merge1}{line:cm_merge2}.
Hence, in every iteration, we construct a set of efficient paths with at most $k$ arcs, possibly dominated by longer paths that have not yet been discovered.

\begin{algorithm}[ht]
\SetKwInOut{Input}{Input} 
\SetKwInOut{Output}{Output}
\Input{Instance $\instance$.}
\Output{Sets of efficient $s,\vrt$-paths $\algeff_\vrt$ for all $\vrt \in \V$.}
\Parameter{Merge function $\mathit{merge}\in \{\maxmerge, \minmerge\}$. }
\BlankLine
$\algeff_{s,0} \leftarrow \{ (s) \}$,
$\algeff_{v,0} \leftarrow \emptyset \, \forall \vrt \in \V\backslash\{s\}$;\\
$k\leftarrow 1$;\\
\While{ $TRUE$ } 
{
\For{$\vrt \in \V$}
{
    $\algeff_{\vrt,k} \leftarrow \algeff_{\vrt,k -1} $;\\
\For{$\vrtalt \in \ingoing{\vrt}$} 
{
    $\algeff_{\vrt,k}.\mathit{merge}(\{ \p + (u,v)   | \p \in \algeff_{\vrtalt,k-1}  \} 
    )$;\\ \label{line:cm_merge2}
} \label{line:cm_merge1}
}
\lIf{$\algeff_{\vrt,k} = \algeff_{\vrt,k-1} \, \forall \vrt\in \V$}
{
    Return $\algeff_{\vrt} := \algeff_{\vrt,k-1} \, \forall \vrt\in\V$
}
{$k \leftarrow k+1$};\\
}
\caption{Label-correcting algorithm for \POSP. \label{algo:corleymoon}}
\end{algorithm}

When $\POSPMIN{}$ shall be solved, we employ a merge procedure that breaks ties, i.e., if 
$\palt$ and $\paalt$ are efficient in $\algeff_{\vrt,k-1} \cup \bigcup_{\vrtalt\in \V} \{ \p + (\vrtalt,\vrt) | \p \in \algeff_{\vrtalt,k-1}  \}$
and $\WW(\palt) = \WW(\paalt)$, either $\palt$ or $\paalt$ is kept but not both.
If $\palt\in\algeff_{\vrt,k-1}$, $\palt$ is kept; if both $\palt$ and $\paalt$ are not in $\algeff_{\vrt,k-1}$, the tie is broken arbitrarily.
We denote this procedure by $\minmerge$.
The procedure $\maxmerge$ only discards a path $\palt$ if it is strictly dominated by another path $\paalt$, i.e.,
$\WW(\paalt) \strictpo \WW(\palt)$. It is used to solve $\POSPMAX{}$ instances.
Both merge procedures can be implemented by pairwise comparison. 
Since efficient paths may contain cycles, \cref{algo:corleymoon} cannot be terminated after $|V|-1$ iterations, as is the case in
the SP and \MOSP{}-variants of Bellman\& Ford's algorithm.

\begin{proposition}\label{theorem:corleymoon_weak_independence}
    Given a  well-posed, his\-to\-ry-free and weakly independent
    instance $\instance$, \cref{algo:corleymoon}$(\minmerge)$ computes a minimal complete set of efficient $s,\vrt$-paths for all $\vrt\in\V$.
\end{proposition}

\begin{proof}
We consider the state of \cref{algo:corleymoon}$(\minmerge)$ after completing iteration $k$.
Assume there is an efficient $s,\vrt$-path $\p = (\vrt_0,\dots,\vrt_k) \in\PS_{s,\vrt}$
of length $k$ with $\p\not\in \algeff_{\vrt,k}$.
Then, there must be an iteration \mbox{$ 0< l < k$} for which
$(\vrt_0,\dots,\vrt_l)\in \algeff_{\vrt_l,l}$ but $(\vrt_0,\dots,\vrt_{l+1}) \not\in \algeff_{\vrt_{l+1},l+1}$.
This means that $(v_0,\dots,v_l)$ was not merged into $\algeff_{\vrt_{l+1},l+1}$
because there was a $\vrt_0,\vrt_{l+1}$-path $\palt \in \algeff_{\vrt_{l+1},l+1}$ 
with $\WW(\palt)\po \WW((\vrt_0,\dots,\vrt_{l+1}))$
but by history-freeness and weak independence, we then have that 
$\WW(\palt + (\vrt_{l+1},\dots,\vrt_k)) = \WW(\p)$.
Now, $\palt + (\vrt_{l+1},\dots,\vrt_k)$ is an efficient $s,\vrt$
-path of the same objective value that survived for at least one iteration longer than $\p$.
Note that $\palt + (\vrt_{l+1},\dots,\vrt_k)$ has a length of at most $|\p| = k$.
Iterating the above argument will yield a path $\paalt$ with $\WW(\p) = \WW(\paalt)$ of length $\tilde{k} \leq k$ with
$\paalt\in\algeff_{\vrt,\tilde{k}} \cap \algeff_{\vrt,k}$ in at most $k-l$ steps.

By $\instance$ being well-posed, $|\bigcup_{\vrt\in V} \nondom_{s,\vrt}|  < \infty$, and hence
there must be some number $M_\vrt > 0$ such that
$\nondom_{s,\vrt} = \WW(\eff_{s,\vrt} \cap \{\p \in \PS_{s,\vrt}: |\p| \leq M_{\vrt} \})$ for all $\vrt\in\V$.
Then, after $M:=\max_{\vrt\in\V} M_\vrt$ iterations, the sets $\algeff_{\vrt, M}$ contain a complete minimal set of efficient $s,\vrt$-paths $\effopt_{s,v}$ for all $\vrt\in\V$.

Now, assume there is some $\vrt\in\V$ such that $\effopt_{s,v}\subsetneq \algeff_{\vrt,M}$, i.e.,
there is a dominated path $\p\in\algeff_{v,M}$.
Since we perform a merge in every iteration, $\p$ cannot be dominated by any path in $\effopt_{s,v}$.
Hence, $\p$ is part of an infinite chain of $s,v$-paths of strictly decreasing weight without a minimum.
This contradicts well-posedness.
Thus, $\algeff_{\vrt,M} = \effopt_{s,v}$.
Then, we must have $\algeff_{\vrt,M} = \algeff_{\vrt,M+1}$ for all $\vrt\in\V$ in iteration $M+1$ and \cref{algo:corleymoon}$(\minmerge)$ terminates.

Finally, assume \cref{algo:corleymoon}$(\minmerge)$ terminates after some iteration $k < M+1$.
Then, $\algeff_{\vrt,k-1} = \algeff_{\vrt,k}$.
Continuing until iteration $M$ gives $\algeff_{\vrt,k} = \algeff_{\vrt,M}$.
Hence, $\algeff_{\vrt} := \algeff_{\vrt,k}$ is a complete minimal set.
\end{proof}

\begin{proposition}\label{theorem:corleymoon_PSWO}
    Given a well-posed, history-free and weakly subpath optimal
    instance $\instance$, \cref{algo:corleymoon}$(\minmerge)$ computes a minimal complete set of efficient $s,\vrt$-paths for all $\vrt\in\V$.
\end{proposition}

\begin{proof}
    The proof follows closely the proof of \cref{theorem:corleymoon_weak_independence}
    and differs only in the initial choice of $\p$ and the construction of $\paalt$.
    A detailed description can be found in \ref{appendix:proof_cm_wso}.
\end{proof}

\begin{rem}
    A curious property of \cref{theorem:corleymoon_PSWO} is that while we use weak subpath optimality to argue correctness, 
    the sets $\algeff_\vrt, \vrt\in\V$ may contain paths that are not subpath optimal.
    This happens, since  \cref{algo:corleymoon} finds for every $\tau\in\nondom_{s,\vrt}$, the path $\p\in\PS_{s,\vrt}$ with
    $\tau = \WW(\p)$ that has the fewest arcs.
\end{rem}

The next theorem formulates the conditions under which \cref{algo:corleymoon} can be applied for $\POSPMAX$.
First, note that any correct $\POSPMAX$\, algorithm can only terminate if $\instance$ is $\efflen$-bounded.

\begin{proposition}\label{theorem:corleymoon_max}
    Consider a $\efflen$-bounded and weakly subpath optimal instance $\instance$.
    For all $\vrt\in\V$,  \cref{algo:corleymoon}$(\maxmerge)$ computes a complete set of efficient $s,\vrt$-paths.
    If $\instance$ is also subpath optimal,  
    it computes a maximal complete set of efficient $s,\vrt$-paths.
\end{proposition}

\begin{proof}

After iteration $k$, consider any efficient $s,\vrt$-path $\p = (\vrt_0,\dots,\vrt_k)$
of length $k$ with $\p\not\in \algeff_{\vrt,k}$.
Then, there must be an iteration $ 0< l < k$ for which
$(\vrt_0,\dots,\vrt_l)\in \algeff_{\vrt_l,l}$ but $(\vrt_0,\dots,\vrt_{l+1}) \not\in \algeff_{\vrt_{l+1},l+1}$.
This means that $(v_0,\dots,v_l)$ was not merged into $\algeff_{\vrt_{l+1},l+1}$
because there was a \mbox{$\vrt_0,\vrt_{l+1}$-path} $\palt \in \algeff_{\vrt_{l+1},l+1}$ 
with $\WW(\palt)\strictpo \WW((\vrt_0,\dots,\vrt_{l+1}))$.
Thus, $\p$ was not subpath optimal.

Let $M \leq \efflen$ be the length of the longest subpath optimal path.
By the above, the algorithm performs at least $M+1$ iterations, and all dominated paths will be removed in the merge-step of the $M+1$-th iteration at the latest.
Since there are no efficient paths of length $> \efflen$, the algorithm terminates after at most $\efflen+1$ iterations.
Hence, $\algeff_\vrt,\vrt\in V$ does not contain dominated paths and is complete, i.e.,  $\WW(\algeff_{\vrt}) = \nondom_{s,\vrt}$.
Now, if $\instance$ is subpath optimal, then no efficient path can be pruned.
Hence, $\algeff_\vrt$ is also a maximal complete set for all $\vrt\in\V$.
\end{proof}

\begin{rem}\label{remark:extending_perny}
     \cite{Perny2005} provide a Bellman-type algorithm for quasi-transitive binary relations on $\SPS$ in acyclic digraphs $\G$.
     In \cref{theorem:corleymoon_max}, $\po$ need no longer be a partial order; antisymmetry and  quasi-transitivity suffice. Hence, \cref{theorem:corleymoon_max} generalizes their result from acyclic to cyclic graphs.
\end{rem}

Despite the intractability of \POSP{},
\cref{thm:corleymoon_complexity} shows that the runtime of \cref{algo:corleymoon}$(\minmerge)$ can be polynomial in the width of $(\VS,\po)$.

\begin{proposition}\label{thm:corleymoon_complexity}
    Let $\instance$ be an instance fulfilling the conditions of  \cref{theorem:corleymoon_weak_independence} or
    \cref{theorem:corleymoon_PSWO}. 
    Furthermore, let $M\in\N$ be the smallest number s.t.  $\nondom_{s,\vrt} = \WW(\eff_{s,\vrt} \cap \{\p \in \PS_{s,\vrt}: |\p| < M \})$ for all $\vrt\in\V$
    and $\wid\in \N\cup\{\infty\}$ the width of $(\VS,\po)$.
    Then,  \cref{algo:corleymoon}$(\minmerge)$ has a worst-case run time of 
    $\ON\left(  \ENC \min \left(  M |\V|^2 \wid^2, |V|^{2M+3} \right) \right).$
\end{proposition}
\begin{proof}
See \ref{appendix:cm}.
\end{proof}

The following example nearly achieves the above bound and
shows that \cref{algo:corleymoon}($\minmerge$) is indeed not output-sensitive.

\begin{example}
Consider the instance $(K_n,\WW,\VS,\po,s)$
where $K_n$ is the complete graph on $n$ vertices,
$\WW: \PS(K_n) \to \VS$ maps into the weight space $\VS :=[n]^M \cup \{ \boldsymbol{0}\}$ for some $M\geq3$, and
\begin{equation}
    \WW\left( \left( \vrt_{i_1},\dots \vrt_{i_l} \right) \right) :=
    \begin{cases}
        (i_1,\dots,i_l) & \text{ if } l \leq M-1 \\
        \boldsymbol{0}  & \text{ if } l > M-1.
    \end{cases}
\end{equation}
The partial order $\po$ on $\VS$ is given by the relation
$\va,\valt \in \VS: \va \po \valt \Leftrightarrow \va = \boldsymbol{0}$.
For $(K_n,\WW,\VS,\po,s)$, history-freeness, well-posedness and weak independence  hold
and \cref{algo:corleymoon}$(\minmerge)$ terminates correctly in iteration \mbox{$M+1$} with every set
$\algeff_{\vrt},\vrt\in\V$ containing exactly one path of length $M$ and weight $\boldsymbol{0}$.
Hence, the output has size $\ON\left(nM\right)$.
However, at the beginning of iteration $M$,
every set $\algeff_{\vrt,M-1},\vrt\in\V$ contains exactly $\sum_{i\in[M-1]} n^i$ paths.
Thus, the \emph{first} merge operation for any $\algeff_{\vrt,M},\vrt\in\V$ has time complexity
%\begin{equation}
%    \Theta\left(\ENC \left(\sum_{i\in[M-1]} n^i\right)^2\right)  = \Theta\left( \ENC n^{2M-2}\right).
%\end{equation}
$\Theta(\ENC (\sum_{i\in[M-1]} n^i)^2)  = \Theta( \ENC n^{2M-2}).$
After the first merge, we have $|\algeff_{\vrt,M}| = 1$, and the time complexity reduces to
\mbox{$\Theta(\sum_{i\in[M-1]} \ENC n^i) = \Theta( \ENC n^{M-1})$}
resulting in an overall time complexity of $\Theta\left( \ENC n^{2M-2}\right)$ for iteration $M$.
For iterations $1,\dots,M-1$, we bound the complexity by
$\Omega ( \ENC  \sum_{k=1}^{M-1} n^2 ((\sum_{i\in[k-1]} n^i)^2) )$ which can be shortened to $\Omega( \ENC n^{2M-2}  )$.
%\begin{equation}
%    \Omega \left( \ENC  \sum_{k=1}^{M-1} n^2 \left(\left(\sum_{i\in[k]} n^i\right)^2\right) \right) 
%    = \Omega  \left( \ENC n^2 n^{2M-2}  \right)  = \Omega \left( \ENC n^{2M}  \right).
%\end{equation}
Hence, the overall time complexity $\Omega \left(\ENC n^{2M-2}  \right)$ is exponential in the output.
\end{example}

\subsection{A Label-Setting Algorithm}\label{subsec:labelsetting}

In the following, we present a  \POSP{} variant of the Multiobjective Dijkstra Algorithm (MDA) \citep{MaristanydelasCasasSedenoNodaBorndoerfer2021}.
However, the arguments put forward easily apply to Martins' algorithm \citep{Martins1984} as well.
The MDA iteratively extracts paths from a priority queue and adds them to a path set $\algeff_\vrt$ for some $\vrt\in \V$.
Such a priority queue sorts its elements by  a total order on $\image(\WW)$.
For \MOSP{}, this is usually the lexicographic order on $\R^d$.
However, \cite{paixao2013} observed that 
for a \MOSP{} instance $(\G,\WW,\R^d_{\geq 0},\leq,s)$ with nonnegative arc weights any binary relation
$\TO$ on $\R^d_{\geq 0}$ fulfilling
\begin{alignat}{3}
     \WW(\p) \po \WW(\palt) \Rightarrow \WW(\p) \TO \WW(\palt)   
    &&  &&                                    \forall \p,\palt \in \PS_{s,t} \, \forall t\in\V  \label{eq:loe}\\
     \WW(\p) \TO \WW(\p + \arc)                    
    &&\quad \quad &&\forall \arc \in \outgoing{t} \,        \forall \p \in \PS_{s,t} \, \forall t\in \V \label{eq:paixao_mon}
\end{alignat} 
can be used instead.
As we will see, the same holds for \POSP{}, i.e.,  any operator $\TO$ fulfilling \cref{eq:loe,eq:paixao_mon} 
can be used to ensure correctness of the MDA for an appropriate \POSP{} instance $\instance$.
Note that \cref{eq:loe} is nothing more than stating that $(\image(\WW),\TO)$ is a linear extension of $(\image(\WW),\po)$.
By the Szpilrajn extension theorem \citep{Szpilrajn1930}, every partial order has at least one linear extension.
However, there are instances in which there is no linear extension respecting  \cref{eq:paixao_mon}.
This is the case, e.g., for \MOSP{} instances with negative arc weights.
For arc-increasing instances, \cref{eq:paixao_mon} is implied by  \cref{eq:loe}.

A linear extension for finite $\PVS$ can be computed in 
$\ON(\tilde{V} + \tilde{A})$ 
where $(\tilde{V},\tilde{A})$ is the acyclic digraph
induced by $\PVS$ \citep{Kahn1962}.
Often, one is also implicitly available, as illustrated in \cref{example:LEO,example:LEO2,example:LEO3}.

\begin{example}\label{example:LEO}
Consider again the instance $(\G,\WW,2^{[n]},\subseteq,s)$ with
$\WW((s)) = \emptyset$ and $\UF(\tau, \arc) := \tau \cup \mathbb{X}_{\arc} \, \forall \tau\in\VS,\arc\in\A$ 
for some $\mathbb{X}\in (2^{[n]})^\A$.
A total order $\TO$ extending $\subseteq$ is given by the \emph{shortlex order}
\begin{equation}
    A \TO B \Longleftrightarrow (|A| < |B|) \lor \left( |A| = |B| \land  \min \left\{ (A \backslash B) \cup (B\backslash A) \right\}
     \in A \right)
\end{equation}
and can be evaluated in $\ON(n)$. %time.
The instance $(\G,\WW,2^{[n]},\subseteq,s)$ is AHW since $(2^{[n]},\cup,\subseteq)$ is a positive partially ordered monoid.
Hence, \cref{eq:paixao_mon} holds.
\end{example}

\begin{example}\label{example:LEO2}
    Consider again the \POSP{}  instance $(\G,\WW,\VS,\po_{\alpha,\beta},s)$ with interval weights from \cref{subsec:interval}.
    The operator $\po_{\alpha,\alpha}$ defines a total order on $\VS$ extending  $\po_{\alpha,\beta}$ for all $-1\leq\alpha\leq\beta\leq -1$ 
    \citep{Okada1994}. Since $(\G,\WW,\VS,\po_{\alpha,\beta},s)$ is AHW, $\po_{\alpha,\beta}$ fulfills \cref{eq:paixao_mon}.
\end{example}

% For electric vehicle routing
\begin{example}\label{example:LEO3}
    Consider an WCSP-R instance $\instance $ as defined in \cref{example:replenishment}.
    The lexicographic order $\po_{lex}$ defined on $\R^2$ by $(\tau_1,\tau_2) \po_{lex} (\tau_1',\tau_2') \Leftrightarrow 
    \tau_1 < \tau_1' \lor (\tau_1 = \tau_1' \land \tau_2\leq\tau_2')$
    is a linear extension of the canonical partial order $\po$ on $\R^2$.
    If every replenishment arc $\arc \in \A'$ has positive cost, this linear extension fulfills \cref{eq:paixao_mon}, even though 
    $\instance$ is not arc-increasing.
    With similar reasoning, we obtain a linear extension fulfilling \cref{eq:paixao_mon} for the EVSP in \cref{example:ELECTRIC}.
\end{example}

We hence assumethe existence of a \emph{linear extension oracle}
\mbox{$\LEO: \VS \times \VS \to \VS$} such that for some total order operator $\TO$ that extends $\po$ it holds that
\mbox{$\LEO(\va,\valt) = \va \Leftrightarrow \va \TO \valt \, \forall \va,\valt \in \V $}. 
The evaluation time of $\LEO$ is assumed to be $\Theta(\ENC)$, i.e., it linearly depends on the problem-specific parameter $\ENC$.

We will now briefly describe \cref{algo:mda}. For an in-depth analysis of its \MOSP{} version, see \cite{MaristanydelasCasasSedenoNodaBorndoerfer2021}.
The algorithm maintains at each vertex $\vrt\in\V$ a set of  permanent efficient walks $\algeff_\vrt$
and a priority queue $H$ of candidate paths sorted by $\LEO$.
The queue $H$ contains at most one $s,\vrt$-path per vertex $\vrt\in\V$, namely the minimal path w.r.t to $\LEO$ among all 
$s,\vrt$-paths that have already been discovered but have not yet been made permanent.
In each iteration, a minimal path, say an \mbox{$s,\vrt$-path} $\p^*$, is chosen from $H$ and made permanent.
It is then propagated  along $\outgoing{\vrt}$.
During propagation, $H$ must be updated to maintain minimality of its entries w.r.t. to $\LEO$.

Finally, a new candidate $s,\vrt$-path is constructed  from $\bigcup_{u\in \ingoing{v}} \algeff_u$ to be inserted into $H$.
By limiting the size of $H$, faster queue operations than in, e.g., Martins' algorithm can be achieved \citep{MaristanydelasCasasSedenoNodaBorndoerfer2021}.
This advantage is paid for by more intricate queue management.

\begin{algorithm}[ht]
    \SetKwInOut{Input}{Input} 
    \SetKwInOut{Output}{Output}
    \Input{Instance $\instance$, linear extension oracle $\LEO$.}
    \Output{Set $\algeff_\vrt$ of efficient paths $\forall{} \vrt \in \V$.}
    \Parameter{Binary Relation $\sqsubset$ (either $\po$ or $\strictpo$).}

        Priority queue $H \leftarrow{} \emptyset{}$; \label{algo:mda:initStart}\\
	\lFor{$\vrt \in \V$}{
	     $\algeff_\vrt \leftarrow{} \emptyset{}$}
        $ H.\mathit{insert}( (s) )$\;
	\BlankLine
	\While{$H \neq \emptyset{}$}{
            
            $\p^* \leftarrow{} H.\mathit{extract\_min_{\LEO}}()$
            \label{line:mda_extraction}\tcp*[r]{A best path w.r.t to $\LEO$.}
	    $\vrt \leftarrow{} \mathit{head}(\p^*)$;\\
	    $\algeff_\vrt.\mathit{append}(\p^*)$\label{algo:mda:linePushBack2};\\\label{line:mda_append}
            \tcc{Find next candidate path for $\vrt$ in $H$.}
            $\p^{new} \leftarrow \argmin_{\LEO} 
            \bigcup_{u\in \ingoing{v}} \bigcup_{{\p \in \algeff_u}}
            \big\{\p_\vrt :=  \p + (u,\vrt) \,| \,\WW(\algeff_\vrt) {\not \sqsubset} \WW(\p_\vrt)
            \land \p_\vrt\notin\algeff_\vrt
            \big\}$\; \label{line:mda_next_candidate}
	    \lIf{$\p^{new} \neq \Null$} {$H.\mathit{insert}(\p^{new})$}
            \tcc{Propagate $\p^{*}$ along $\outgoing{v}$ into $H$. }
            \For{$u \in \outgoing{\vrt}$}
                {
                    $\p \leftarrow \p^{*} + (v,u)$;\\
                \uIf{$\WW(\algeff_{u}) \not\sqsubset \WW(\p)$  }
                { \label{line:mda_dom_check}
                    \lIf{$! H.\mathit{contains}(u)$}
                    {$H.\mathit{insert}(\p)$}
                    \lElseIf{$\LEO \left( \p,H.\mathit{getPath}(u)\right) = \p$}
                    {$H.\mathit{update}(\p)$} \label{line:mda_propagate_end}
                }

                } \label{line:mda_propagate_start}
	}
	\Return $\algeff_\vrt \, \forall \vrt\in \V$; 
        \caption[caption]{\POSP{}-variant of the MDA\label{algo:mda}.}
\end{algorithm}

Note that dominance checks can be performed using one of the binary relations 
$\strictpo$ and $\po$ depending on whether one aims to solve $\POSPMAX{}$ or $\POSPMIN{}$.
The corresponding versions of the algorithm are denoted by \cref{algo:mda}($\strictpo$) and  \cref{algo:mda}($\po$), respectively.

In the following, we show correctness of \cref{algo:mda}. The proof is based on the following lemma
mirroring a similar result by \cite[Thm. 4.1]{paixao2013} on using alternative total orders in Martins' algorithm for \MOSP{}.

\begin{lemma}\label{lemma:MDA_PERMANENT_EFFICIENT}
    Given a  linear extension oracle $\LEO$ fulfilling \cref{eq:paixao_mon}
    for a well-posed and weakly subpath optimal instance \mbox{$\instance$}, every  path $\p^*$ made permanent in \cref{line:mda_append} of \cref{algo:mda}$(\strictpo)$ is subpath optimal.
    If \mbox{$\instance$} is also history-free, every  path $\p^*$ made permanent in \cref{line:mda_append} of \cref{algo:mda}$(\po)$ is subpath optimal.
\end{lemma}

\begin{proof}
We first consider \cref{algo:mda}$(\po)$.
Assume that in iteration $k$,  for the first time, an  $s,v$-path $\p\in\PS_{s,\vrt}$ selected in \cref{line:mda_append} is dominated, i.e., there is an efficient and subpath optimal path $\palt \in \PS_{s,\vrt}$ with  $\WW(\palt) \strictpo \WW(\p)$.

Consider the iteration $l<k$ in which $\p$ was added to $H$ for the last time.
By the dominance checks in \cref{line:mda_next_candidate,line:mda_dom_check}, $\WW(\palt) \not \in\WW(\algeff_\vrt)$.
Between iterations $l$ and $k$, $\p$ remains in $H$. Hence, $\palt\not\in\algeff_\vrt$ in \cref{line:mda_extraction}
of iteration $k$.
Let $\palt=(s= \vrt_1,\dots,\vrt_n = \vrt)$ and
$i\in [n]$ the maximal integer such that $(\vrt_1,\dots,\vrt_i)\in \algeff_{\vrt_i}$ in \cref{line:mda_extraction} of iteration $k$.
Now, if $(\vrt_1,\dots,\vrt_{i+1})\in H$, we obtain a contradiction by \cref{eq:paixao_mon} as $\WW((\vrt_1,\dots,\vrt_{i+1}))\TO\WW(\palt)\strictto \WW(\p)$.
If $(\vrt_1,\dots,\vrt_{i+1}) \not \in H$,
$(\vrt_1,\dots,\vrt_{i+1})$ might have been replaced in $H$ by $\paalt$ with $\WW(\paalt)\po\WW((\vrt_1,\dots,\vrt_{i+1}))$
resulting again in the same contradiction.
If neither is the case, $(\vrt_1,\dots,\vrt_{i+1})$ must have  been rejected either in \cref{line:mda_next_candidate}
or \cref{line:mda_dom_check} by $\WW(\algeff_{\vrt_{i+1}}) \po \WW((v_1,\dots,v_{i+1}))$.
Hence, there is $\paalt\in \algeff_{\vrt_{i+1}}$ with $\WW(\paalt) = \WW((v_1,\dots,v_{i+1}))$.
By history-freeness, 
$\WW( \paalt + (\vrt_{i+1},\dots,\vrt_{n})) = \WW(\palt)$. 
We can now repeat the same argument with $\paalt$ instead of $\palt$. Doing so, we will eventually find
the contradiction.

For \cref{algo:mda}$(\po)$, this third case can never happen. Hence, we obtain a contradiction without requiring history-freeness of $\instance$.

Thus, all paths made permanent are efficient. 
By only building new paths from permanent paths, this implies all permanent paths to be subpath optimal.
\end{proof}

\begin{proposition}\label{thm:MDA_MIN}
    Given a  well-posed, history-free and weakly subpath optimal instance \mbox{$\instance$}
    and a linear extension oracle $\LEO$ fulfilling \cref{eq:paixao_mon},
    \mbox{\cref{algo:mda}}$(\po)$ computes a minimal complete set of efficient $s,\vrt$-paths for all $\vrt\in\V$.
\end{proposition}

\begin{proof}
     \Cref{lemma:MDA_PERMANENT_EFFICIENT} ensures that every path set $\algeff_{\vrt}$ contains only efficient paths at any point during the algorithm.
    % Termination
    Since \mbox{$|\bigcup_{\vrt\in \V} \nondom_{s,\vrt}|$} is finite by $\instance$ well-posed, 
    there must be a point at which no $\p^{new}$ can be found with $\WW(\algeff_{\vrt}) \npo \WW(\p^{new}$).
    At that point, no new paths are added to $H$, and hence, eventually $H = \emptyset$ will hold.
    Thus, \cref{algo:mda} $(\po)$ terminates.
    % Minimality
    The minimality of all path sets $\algeff_\vrt,\vrt\in \V$ is ensured by choosing $\p^{new}$ such that  $\WW(\algeff_{\vrt}) \npo \WW(\p^{new})$ in \cref{line:mda_next_candidate}.
    Finally, completeness can be shown similarly to previous proofs by using history-freeness.
    %Finally, it remains to show that all label sets are complete when $|H|  = \emptyset$.
    %Assume that there is some $\tau \in\nondom_{s,\vrt} \backslash \WW(\algeff_{\vrt})$ for some $\vrt\in\V$. 
    %Then, there must be some subpath optimal path $\p =(s=u_0,\dots,u_k=\vrt) \in\PS_{s,\vrt}$ with $\WW(\p) =\tau$
    %containing a subpath $(u_0,\dots,u_j)$ with 
\end{proof}

\begin{rem}
    In contrast to  \cref{algo:corleymoon},  \cref{algo:mda}
    propagates only paths  that are known to be efficient.
    Hence, there is no label-setting equivalent of \cref{theorem:corleymoon_weak_independence} relying on weak independence.
\end{rem}

\begin{proposition}\label{thm:MDA_MAX}
    Consider a  $\efflen$-bounded, weakly subpath optimal instance \mbox{$\instance$}
    and any linear extension oracle $\LEO$ fulfilling \cref{eq:paixao_mon}.
    Then, \mbox{\cref{algo:mda}}$(\strictpo)$ computes a complete set of efficient $s,\vrt$-paths for all $\vrt\in\V$.
    If $\instance$ is also subpath optimal, \mbox{\cref{algo:mda}}$(\strictpo)$ computes a maximal complete set of efficient $s,\vrt$-paths for all $\vrt\in\V$.
\end{proposition}

\begin{proof}
    Remember that $\efflen$-bounded \POSP{} instances are well-posed.
    Therefore, \cref{lemma:MDA_PERMANENT_EFFICIENT} applies, and every path in $\algeff_\vrt,\vrt\in\V$ is efficient and subpath optimal.
    Since $\instance$ is $\efflen$-bounded, there must be a point at which no new $\p^{new}$ can be found with $\WW(\algeff_{\vrt}) \strictNpo \WW(\p^{new})$.
    At that point, no new paths are added to $H$, and hence, eventually $H=\emptyset$ will hold.
    Completeness is ensured by choosing $\p^{new}$ such that $\WW(\algeff_{\vrt} ) \strictNpo \WW(\p^{new})$.
    If $\instance$ is subpath optimal, no efficient paths can be pruned. Hence, $\algeff_\vrt$ must be maximal for all $\vrt\in\V$ at
    termination.
\end{proof}

For arc-increasing \POSP{} instances, \cref{eq:paixao_mon} follows from \cref{eq:loe} for any linear extension of $\po$.
Hence, we formulate the following corollaries.

\begin{corollary}\label{col:col1}
    Given an AHW instance $\instance$ and any linear extension oracle $\LEO$, \mbox{\cref{algo:mda}} $(\po)$ computes a minimal complete set of efficient $s,\vrt$-paths for all $\vrt\in\V$.
\end{corollary}

\begin{table}[t!]
    \small
    \setlength{\tabcolsep}{2pt}
    \centering
    \begin{tabular}{llcccccccccccl}
         \toprule
         Algorithm &  Problem     & WP  & H   & CND & CI & A   & WI  & I   & WSO & SO  & $\efflen$  & $21$ & Prop.    \\
         \midrule                                                            
         Bellman   & $\POSPMIN{}$ & \fb & \fb &      &     &     & \fb &     &     &     &     &     &  \ref{theorem:corleymoon_weak_independence}             \\    % Thm 5.1
         Bellman   & $\POSPMIN{}$ & \fb & \fb &      &     &     &     &     & \fb &     &     &     & \ref{theorem:corleymoon_PSWO}                          \\    % Thm 5.2
         Bellman   & $\POSPMIN{}$ & \ob & \fb &      &     &     &     &     & \fb &     & \fb &     & \ref{theorem:corleymoon_PSWO}                          \\
         Bellman   & $\POSPMIN{}$ & \ob & \fb & \ob  & \fb &     & \fb &     & \ob &     &     &     &  \ref{thm:WIPSO},  \ref{theorem:corleymoon_PSWO}        \\                  
         Bellman   & $\POSPMIN{}$ & \ob & \fb & \ob  & \ob & \fb & \fb &     & \ob &     &     &     & \ref{thm:WIPSO},  \ref{theorem:corleymoon_PSWO}        \\                  
         % This case is already covered by the first one. No application of 5.2 + 4.5 needed
         %Bellman   & $\POSPMIN{}$ & \fb & \fb & \fb  &     &     & \fb &     & \ob &     &     &     & Thm. \ref{thm:WIPSO3}, \ref{theorem:corleymoon_PSWO}        \\
         Bellman   & $\POSPMIN{}$ & \ob & \fb &      &     &     & \fb &     & \ob &     & \fb &     &  \ref{thm:WIPSO2}, \ref{theorem:corleymoon_PSWO}        \\
         Bellman   & $\POSP{}$    & \ob &     &      &     &     &     &     & \fb &     & \fb &     &  \ref{theorem:corleymoon_max}                           \\
         % Already shown by the row with same conditions for MIN variant
         %Bellman   & $\POSP{}$    & \ob & \fb &      &     &     & \fb &     & \ob &     & \fb &     & Thm. \ref{thm:WIPSO2}, \ref{theorem:corleymoon_max}         \\
         Bellman   & $\POSPMAX{}$ & \ob &     &      &     &     &     &     & \ob & \fb & \fb &     &  \ref{theorem:corleymoon_max}                           \\
         Bellman   & $\POSPMAX{}$ & \ob &     &      &     &     &     & \fb & \ob & \ob & \fb &     &  \ref{thm:ISWO}, \ref{theorem:corleymoon_max}           \\
         MDA       & $\POSPMIN{}$ & \fb & \fb &      &     &     &     &     & \fb &     &     & \fb &  \ref{thm:MDA_MIN}                                      \\           
         MDA       & $\POSPMIN{}$ & \fb & \fb & \ob  & \ob & \fb &     &     & \fb &     &     & \ob &  \ref{thm:MDA_MIN}                                      \\
         MDA       & $\POSPMIN{}$ & \ob & \fb &      &     &     &     &     & \fb &     & \fb & \fb &  \ref{thm:MDA_MIN}                                      \\           
         MDA       & $\POSPMIN{}$ & \ob & \fb & \ob  & \ob & \fb &     &     & \fb &     & \fb & \ob &  \ref{thm:MDA_MIN}                                      \\           
         MDA       & $\POSPMIN{}$ & \ob & \fb & \ob  & \fb &     & \fb &     & \ob &     &     & \fb &  \ref{thm:WIPSO}, \ref{thm:MDA_MIN}                     \\
         MDA       & $\POSPMIN{}$ & \ob & \fb & \ob  & \ob & \fb & \fb &     & \ob &     &     & \ob &  \ref{thm:WIPSO}, \ref{thm:MDA_MIN}                     \\
         MDA       & $\POSPMIN{}$ & \ob & \fb &      &     &     & \fb &     & \ob &     & \fb & \fb &  \ref{thm:WIPSO2}, \ref{thm:MDA_MIN}                    \\
         MDA       & $\POSPMIN{}$ & \fb & \fb & \fb  &     &     & \fb &     & \ob &     &     & \fb &  \ref{thm:WIPSO3}, \ref{thm:MDA_MIN}                    \\
         MDA       & $\POSP{}$    & \ob &     &      &     &     &     &     & \fb &     & \fb & \fb &  \ref{thm:MDA_MAX}                                      \\
         MDA       & $\POSP{}$    & \ob &     & \ob  & \ob & \fb &     &     & \fb &     & \fb & \ob &  \ref{thm:MDA_MAX}                                      \\
         MDA       & $\POSP{}$    & \ob & \fb &      &     &     & \fb &     & \ob &     & \fb & \ob &  \ref{thm:WIPSO2}, \ref{thm:MDA_MAX}                    \\
         MDA       & $\POSPMAX{}$ & \ob &     &      &     &     &     &     & \ob & \fb & \fb & \fb &  \ref{thm:MDA_MAX}                                      \\             
         MDA       & $\POSPMAX{}$ & \ob &     &      &     &     &     & \fb & \ob & \ob & \fb & \fb &  \ref{thm:ISWO}, \ref{thm:MDA_MAX}                      \\             
         MDA       & $\POSPMAX{}$ & \ob &     & \ob  & \ob & \fb &     &     & \ob & \fb & \fb & \ob &  \ref{thm:MDA_MAX}                                      \\             
         MDA       & $\POSPMAX{}$ & \ob &     & \ob  & \ob & \fb &     & \fb & \ob & \ob & \fb & \ob &  \ref{thm:ISWO}, \ref{thm:MDA_MAX}                      \\             
        \bottomrule
    \end{tabular}
    \caption{
    Summary of the results of \cref{sec:dpp,sec:algorithms} giving an overview of which algorithm can be used for which \POSP{} variant and weight structure.
    The possible optimality conditions are  well-posedness (WP), history-freeness (H),  cycle-non-decreasingness (CND), cycle-increasingness (CI),  arc-increasingness (A), weak independence (WI), independence (I),
    weak subpath optimality (WSO), subpath optimality (SO), $\efflen$-boundedness ($\efflen$), and \cref{eq:paixao_mon} (19).
    Bellman refers to \cref{algo:corleymoon}; MDA refers to \cref{algo:mda}.
    Full bullets \fb indicate requirements for correctness; empty bullets \ob indicate properties that are implied by the specific combination of required optimality conditions.
    Several rows are special cases of others but should be easier to verify than the general case.
    }
    \label{tab:summary}
\end{table}

\begin{corollary}\label{col:col2}
    Given an arc-increasing, $\efflen$-bounded, weakly subpath optimal instance $\instance$ and any linear extension oracle $\LEO$,
    \mbox{\cref{algo:mda}} $(\strictpo)$ computes a complete set of efficient $s,\vrt$-paths for all $\vrt\in\V$.
    If $\instance$ is also subpath optimal, \mbox{\cref{algo:mda}}$(\strictpo)$ computes a maximal complete set of efficient $s,\vrt$-paths for all $\vrt\in\V$.
\end{corollary}
Letting $N_{max}:= \max_{\vrt\in\V}|\algeff_{\vrt}|$,
the time complexity of \cref{algo:mda} is 
$\ON( \ENC N_{max}  ( |\V| \log(|\V|) + N_{max}|\A| ) )$
for both its $\POSPMIN{}$ and $\POSPMAX{}$ variants if \cref{line:mda_next_candidate} is implemented efficiently.
The proof follows to the letter the one given by \cite{MaristanydelasCasasSedenoNodaBorndoerfer2021} 
for the \MOSP{} case and will not be repeated here.
Hence, \cref{algo:mda} is output-sensitive.

Finally, we combine the findings of this section with \cref{sec:dpp}  in  \cref{tab:summary}.
It reports, for every problem variant of \POSP{}, the conditions that 
must be fulfilled to ensure correctness of \cref{algo:corleymoon,algo:mda}, allowing to quickly identify algorithms for \POSP{} problems
as seen in \cref{example:MartinsToMDA,example:LabelSettingWCSPR,example:LabelSettingElectric}.

\begin{example}\label{example:MartinsToMDA}
    Following \cref{tab:summary}, the MDA can be used for all instances that are AHW. Examples are given in \cref{subsec:motdsp,subsec:bottleneck,subsec:pomonoid,subsec:interval}.
    Moreover, the currently best known running time bound for Martins' algorithm is  $\ON\left(\ENC N_{max}^2 |V|^3 \right)$ \citep{MaristanydelasCasasBorndoerferKrausetal.2021}.
    We hence immediately obtain an improvement over the time complexity that can be achieved with the variants of Martins' algorithm proposed in
    \cite{Okada1994, Okada2000, Gandibleux2006, Vu2022}.
\end{example}

\begin{example}\label{example:LabelSettingElectric}
    In \cref{example:ELECTRIC}, we've seen the electric vehicle shortest path problem with recuperation and recharging  to be 
    well-posed, history-free and weakly subwalk optimal. Following \cref{example:LEO3}, there is a linear extension fulfilling \cref{eq:paixao_mon}.
    Hence, a cost minimal feasible path can be computed using the label-setting \cref{algo:mda}$(\po)$.
\end{example}

\begin{example}\label{example:LabelSettingWCSPR}
    The weight-constrained shortest path problem with replenishment $(\G,c,\VS,\po,s)$, 
    as defined in \cref{example:replenishment}, 
    is well-posed, history-free and weakly independent.
    We can hence apply the label-correcting \mbox{\cref{algo:corleymoon}$(\minmerge)$} to find minimal sets of efficient paths.
    If every replenishment arc $a\in A'$ has positive cost $w(a) >0$, the WCSP-R instance is weakly subpath optimal, and, by \cref{example:LEO3}, there is linear extension of $\po$  fulfilling \cref{eq:paixao_mon}.
    It can be hence solved using the MDA.
    To the best of our knowledge, only a label-correcting algorithm was previously known in the literature \citep{Smith2012}.
\end{example}

\section{Conclusion}\label{sec:conclusion}

We introduced \POSP{}, a generalization of \MOSP{} to general partial orders.
Many variants and extensions of \MOSP{} in the literature are special cases of \POSP{}.
These extensions have  been studied independently of each other and are usually solved
with ad-hoc adaptations of Martins' algorithm. We have found that they rely on the same  structural properties, namely 
arc-increasingness, history-freeness, and weak independence and can hence be addressed using label-setting and label-correcting \MOSP{} algorithms.

As a byproduct, we have shown that all these variants can be solved with the recently introduced MDA without having to reexamine each variant individually. 
This is particularly meaningful as the best known output-sensitive running time bound for the MDA is better than the best known bound for Martins' algorithm.

Finally, we provided a taxonomy of the \POSP{} variants for which labeling methods can be correctly applied 
to either their maximization or minimization variants. We hope that this overview that was missing in the literature helps researchers and practitioners alike to decide whether a labeling technique is suitable for their custom \POSP{} variant.

While the optimality conditions derived in this paper are easy to check for all known applications, it is still an application-dependent task.
Whether a general, efficient procedure to check them can be found is open.

%\hbadness=6000
%\printbibliography
\bibliographystyle{apalike-ejor}
\bibliography{lit.bib}

\appendix
\section{AHW'ness for \cref{subsec:interval}}\label{appendix:okada}
%For any  $\va,\valt,\vao\in\VS$ 
%with $\va \strictpo_{\alpha,\beta} \valt$, it holds
%\begin{equation}
%    \begin{aligned}
%            & \va \strictpo_{\alpha,\beta} \valt \\
%          \implies & \gamma(\va_w-\valt_w)   <  \valt_c - \va_c                   %           & \text{for } \gamma\in\{\alpha,\beta \} \\
%   \implies & \gamma(\va_w-\valt_w + \vao_w-\vao_w)   <  \valt_c - \va_c +\vao_c %- \vao_c  & \text{for } \gamma\in\{\alpha,\beta \} \\
%   \implies & \va + \vao \strictpo_{\alpha,\beta} \valt + \vao.\\
%    \end{aligned}
%\end{equation}
For any  $\va,\valt,\vao\in\VS$ 
with $\va \strictpo_{\alpha,\beta} \valt$, we have
$\gamma(\va_w-\valt_w)   <  \valt_c - \va_c $ for $\gamma\in\{\alpha,\beta \}$.
Adding zero gives
$\gamma(\va_w-\valt_w + \vao_w-\vao_w)   <  \valt_c - \va_c +\vao_c - \vao_c$
which implies $\va + \vao \strictpo_{\alpha,\beta} \valt + \vao$.
%
%\begin{equation}
%    \begin{aligned}
%            & \va \strictpo_{\alpha,\beta} \valt \\
%          \implies & \gamma(\va_w-\valt_w)   <  \valt_c - \va_c                   %           & \text{for } \gamma\in\{\alpha,\beta \} \\
%   \implies & \gamma(\va_w-\valt_w + \vao_w-\vao_w)   <  \valt_c - \va_c +\vao_c %- \vao_c  & \text{for } \gamma\in\{\alpha,\beta \} \\
%   \implies & \va + \vao \strictpo_{\alpha,\beta} \valt + \vao.\\
%    \end{aligned}
%\end{equation}
%
Thus, independence and subpath optimality follow.
Finally, arc-increasingness of  $(\G,\WW,\VS,\po_{\alpha,\beta},s)$ holds since for all $\va,\valt\in\VS$ and $-1\leq \alpha \leq \beta\leq 1$ and $\gamma \in \{\alpha, \beta\}$, we have
\begin{equation*}
    %\begin{aligned}
        \va  \po_{\alpha,\beta} \va + \valt  
        \iff  \gamma(\va_w-\va_w - \valt_w)  \leq  \va_c + \valt_c - \va_c
        %
        %\iff  -\gamma\valt_w \leq   \valt_c           
        %&& \text{ for } \gamma\in\{\alpha,\beta \} \\
        \iff  -\gamma \leq \valt_c / \valt_w.
    %\end{aligned}
\end{equation*}

\section{Proof of \cref{theorem:corleymoon_PSWO}}\label{appendix:proof_cm_wso}

 Let $\tau\in \nondom_{s,\vrt}$.
 Then, by weak subpath optimality, there is an efficient, subpath optimal path $\p \in\PS_{s,\vrt}$
 such that $\WW(\p) = \tau$.
    Let $\p  = (\vrt_0,\dots,\vrt_k)$ and consider the state of  \cref{algo:corleymoon} after the $k$'th iteration.
    Assume $\p\not\in \algeff_{\vrt,k}$.
    Then, there must be an iteration $ 0< l < k$ for which
    \mbox{$(\vrt_0,\dots,\vrt_l)\in \algeff_{\vrt_l,l}$} but $(\vrt_0,\dots,\vrt_{l+1}) \not\in \algeff_{\vrt_{l+1},l+1}$.
    This means the path $(v_0,\dots,v_l)$ was not merged into the set $\algeff_{\vrt_{l+1},l+1}$ in iteration $l+1$
    because there was already $\vrt_0,\vrt_{l+1}$-path $\palt \in \algeff_{\vrt_{l+1},l+1}$ 
    with \mbox{$\WW(\palt)\po \WW((\vrt_0,\dots,\vrt_{l+1}))$}.
    Since $(\vrt_0,\dots,\vrt_{l+1})$ is efficient, it must be that 
    $\WW(\palt) =  \WW((\vrt_0,\dots,\vrt_{l+1}))$.
    Hence, by history-freeness,
    \begin{equation}\label{eq:corley_proof_2}
    \WW(\palt + (\vrt_{l+1},\dots,\vrt_j)) = \WW((\vrt_0,\dots,\vrt_j)) \, \forall j \in \{l+2,\dots,k\}.
    \end{equation}
    Now, $\palt + (\vrt_{l+1},\dots,\vrt_k)$ is of length at most $k$ and survives at least until iteration $l+1$.
    By \cref{eq:corley_proof_2}, 
    we can repeat the above procedure at most $k-l$ times to find a path $\paalt\in\PS_{s,\vrt}$ with $\WW(\paalt) = \WW(\p)$ and
    $\paalt \in \algeff_{\vrt,k}$.
    Note that neither $\paalt$ nor any intermediate path are necessarily subpath optimal.  
    The rest of the proof is identical to the one of \cref{theorem:corleymoon_weak_independence}.

\section{Proof of \cref{thm:corleymoon_complexity}}\label{appendix:cm}
\begin{proof}
    Note that $M$ is the length of the longest path in the solution set and
    $M+1$ the number of iterations performed by \cref{algo:corleymoon}$(\minmerge)$. 
    The time complexity is dominated by the complexity of the merge operation in \cref{line:cm_merge2}.
    In iteration $k<M+1$, any merge requires up to $|\algeff_{\vrt,k}|+|\algeff_{u,k-1}|$ evaluations of paths and  $|\algeff_{\vrt,k}| |\algeff_{u,k-1}|$ comparisons.
    Both operations take $\Theta(\ENC)$ time.
    We have $|\algeff_{\vrt,k}| \leq \min \left( \wid, \sum_{i=0}^k |\V|^i \right)$. 
    The bound is tight if $\G$ is the complete graph and all paths are incomparable.
    Hence, we have
    \begin{equation}
     \begin{aligned}
        &|\algeff_{\vrt,k}| |\algeff_{u,k-1}| \leq   \min \left( \wid, \sum_{i=0}^k |\V|^i \right)  \min \left( \wid, \sum_{i=0}^{k-1} |\V|^i \right)  \\ 
        &= \min \left( \wid, \frac{|V|^{k+1} -1}{|V|-1} \right)  \min \left( \wid, \frac{|V|^{k} -1}{|V|-1} \right) 
        \in \ON \left( \min \left(  \wid^2, |V|^{2k-1}   \right) \right)
     \end{aligned}        
    \end{equation}
    and a merge in iteration $k$ can be done in
    $\ON \left( \ENC \min \left(  \wid^2, |V|^{2k-1}   \right) \right).$
   % \begin{alignat}{3}
   %     \ON \left( \ENC \min \left(  \wid^2, |V|^{2k-1}   \right) \right).
   % \end{alignat}
    Since there are $|A|$ $\mathit{merge}$-operations per iteration, we obtain a time complexity of 
    \begin{equation}
    \begin{aligned}
    \ON\Bigg(\ENC \sum_{k=1}^{M+1} |V|^2 \min \Big(  \wid^2, |V|^{2k-1} \Big)  \Bigg)  
    & \subseteq  \ON  \left( \ENC \min \left(M |V|^2 \wid^2,  \sum_{k=1}^{M+1} |V|^{2k+1}   \right) \right) \\
         & = \ON  \left( \ENC \min \left(M |V|^2 \wid^2,  |V|^{2M+3}   \right) \right). 
     \end{aligned}\qedhere
    \end{equation}
\end{proof}

\end{document}